        \documentclass[11pt,a4paper]{article}
        \usepackage{jcappub}
	\usepackage{indentfirst}
        \usepackage{acronym}
  	\usepackage{appendix}
        \usepackage{amsmath}
        \usepackage{amssymb}
        \usepackage{lscape}
        \usepackage{wasysym} 
        \usepackage{verbatim}
        \usepackage{graphicx}
        \usepackage{hyperref}
        \usepackage{color}
\def\lsim{\,\lower2truept\hbox{${< \atop\hbox{\raise4truept\hbox{$\sim$}}}$}\,}
\def\gsim{\,\lower2truept\hbox{${> \atop\hbox{\raise4truept\hbox{$\sim$}}}$}\,}

\def\deg{\ifmmode^\circ \else$^\circ $\fi}    
\def\arcs{\ifmmode {'' }\else $'' $\fi}     
\def\arcm{\ifmmode {' }\else $' $\fi}     
\def\buildrel#1\over#2{\mathrel{\mathop{\null#2}\limits^{#1}}}
\def\mper{\ifmmode \buildrel m\over . \else $\buildrel m\over .$\fi}
\def\hper{\ifmmode \rlap.^{h}\else $\rlap{.}^h$\fi}
\def\sper{\ifmmode \rlap.^{s}\else $\rlap{.}^s$\fi}
\def\arcsper{\ifmmode \rlap.{' }\else $\rlap{.}' $\fi}
\def\arcmper{\ifmmode \rlap.{'' }\else $\rlap{.}'' $\fi}

\title{CMB polarization anisotropies from cosmological reionization: extension to B-modes}
\author[a,b]{T. Trombetti,} \author[a,c]{C. Burigana}
\affiliation[a]{INAF-IASF Bologna,
Via P. Gobetti 101, I-40129, Bologna, Italy} \affiliation[b]{Dipartimento di Fisica, Universit\`a La Sapienza, 
P.le A. Moro 2, I-00185 Roma, Italy} \affiliation[c]{Dipartimento di Fisica, Universit\`a degli Studi di Ferrara, 
Via G. Saragat 1, I-44100 Ferrara, Italy}
\emailAdd{trombetti@iasfbo.inaf.it} \emailAdd{burigana@iasfbo.inaf.it}
\abstract{The accurate understanding of the ionization history of the Universe plays a fundamental role in modern cosmology. It includes a phase of cosmological reionization after the standard recombination epoch, 
possibly associated to the early stages of structure and star formation. 
While the simple ``$\tau$-parametrization'' of the reionization process and, in particular,
of its imprints on the cosmic microwave background (\acs{CMB}) anisotropy likely represents a sufficiently accurate modelling
for the interpretation of current \acs{CMB} data, a great attention has been recently posed on the 
accurate computation of the reionization signatures in the \acs{CMB} for
a large variety of astrophysical scenarios and physical processes.
This work is aimed at a careful characterization of the imprints introduced in the polarization anisotropy, with particular attention to the B-modes.
We have implemented a modified version of \acs{CAMB}, the Cosmological Boltzmann code for computing the angular power spectrum (\acs{APS}) of the anisotropies of the \acs{CMB}, 
to introduce the hydrogen and helium ionization fractions predicted in astrophysical and phenomenological reionization histories, beyond the simple $\tau$-parametrization.
We compared the results obtained for these models for all the non-vanishing (in the assumed scenarios) modes of the \acs{CMB} \acs{APS}.
The amplitude and shape of the B-mode \acs{APS} depends, in particular, on the tensor-to-scalar ratio, $r$, and on the reionization history, thus an accurate modeling of the reionization process will have implications for the precise determination 
of $r$ or to set more precise constraints on it through the joint analysis of E and B-mode polarization data available in the next future and from a mission of next generation. Considering also the limitation from potential residuals of astrophysical foregrounds, we discussed the capability of next data to disentangle between different reionization scenarios in a wide range of tensor-to-scalar ratios.}
\begin{document} \maketitle

\flushbottom

\section*{List of acronyms}
\begin{acronym}[WMAP]
\acsfont
\acro{CMB}[CMB]{Cosmic Microwave Background}
\acro{COrE}[COrE]{Cosmic Origins Explorer}
\acro{WMAP}[WMAP]{Wilkinson Microwave Anisotropy Probe}
\acro{IGM}[IGM]{Inter Galactic Medium}
\acro{SDSS}[SDSS]{Sloan Digital Sky Survey}
\acro{QSO}[QSO]{Quasar}
\acro{IMF}[IMF]{Initial Mass Function}
\acro{CAMB}[CAMB]{Code for the Anisotropy Microwave Background}
\acro{APS}[APS]{Angular Power Spectrum}
\acro{CF06}[CF06]{Suppression Reionization Model} 
\acro{G00}[G00]{Filtering Reionization Model}
\acro{CV}[CV]{Cosmic Variance}
\acro{SV}[SV]{Sampling Variance}
\acro{IR}[IR]{InfraRed}
\acro{ACT}[ACT]{Atacama Cosmology Telescope}
\acro{ALMA}[ALMA]{Atacama Large Millimeter/submillimeter Array}
\acro{FWHM}[FWHM]{Full Width Half Maximum}
\acro{HFI}[HFI]{HFI Frequency Instrument}
\acro{LFI}[LFI]{Low Frequency Instrument}
\acro{CDM}[CDM]{Cold Dark Matter}

\end{acronym}
\section{Introduction}

The accurate understanding of the ionization history of the Universe plays a fundamental role in modern cosmology.
The classical theory of hydrogen recombination for pure baryonic cosmological models \cite{peebles, zeldovich} has been
subsequently extended to non-baryonic dark matter models \cite{zabotin, jones, recfast} and
recently accurately updated to include also helium recombination in the current cosmological scenario 
(see e.g. \cite{switzer_hirata_07_3} and references therein).
Various models of the subsequent Universe ionization history have been considered to take into account
additional sources of photon and energy production, possibly associated to the early stages of structure and star formation, able to significantly
increase the free electron fraction, $x_e$, above the residual fraction ($\sim10^{-3}$) after the standard recombination epoch
at $z_{\rm rec} \simeq 10^3$. 
These photon and energy production processes associated to the cosmological reionization phase may leave imprints in the \acfi{CMB} providing a crucial 
``integrated'' information on the so-called {\it dark} and {\it dawn} {\it ages}, i.e. the epochs before or at the beginning the formation of first cosmological structures.
For this reason, among the extraordinary results achieved by the \acfi{WMAP} mission\footnote{http://lambda.gsfc.nasa.gov/product/map/current/},
the contribution to the understanding of the cosmological reionization process has received a great attention.

This work is aimed at a careful characterization of the imprints introduced in the polarization anisotropy, with particular attention to the B-modes.

In Sect. \ref{cosmoreion} we briefly summarize the current observational information coming from \acs{WMAP} on the cosmological reionization and describe its main imprints in the \acs{CMB}.
In Sects. \ref{astromodel} and \ref{phenmodel} we discuss the main properties of the two classes of reionization models considered in this work, astrophysical and phenomenological, respectively.
Sect. \ref{code} concerns the numerical implementation we carried out to include the considered reionization scenarios in our Boltzmann code modified version. 
The experimental sensitivity of on-going and future \acs{CMB} anisotropy space missions and the limitation coming from astrophysical foregrounds are discussed in Sect. \ref{limitations}, 
while our main results are presented in Sect. \ref{results}. Two appendices report on some technical details of this work. Finally, in Sect. \ref{conclu} we draw our conclusions.

\section{Cosmological reionization}
\label{cosmoreion}

To first approximation, the beginning of the reionization process is identified
by the Thomson optical depth,
$\tau$. The values of $\tau$ compatible with \acs{WMAP} 3yr data, possibly complemented 
with external data, are typically in the range $\sim 0.06-0.12$ (corresponding to
a reionization redshift in the range $\sim 8.5-13.5$ for a sudden reionization 
history), the exact interval depending on the
considered cosmological model and combination of data sets \cite{Spergel2007}. Subsequent  \acs{WMAP} data releases 
improved the measure of $\tau$, achieving a 68\% uncertainty of $\simeq \pm 0.015$ \cite{dunkley2009,komatsu2009,larson2011,komatsu2011}. 
Under various hypotheses (simple $\Lambda$CDM model with six parameters, 
inclusion of curvature and dark energy, of different kinds of isocurvature modes, of neutrino properties, of primordial helium mass fraction, 
or of a reionization width) the best fit of $\tau$ lies in the range $\simeq 0.086-0.089$, while allowing for the presence of primordial tensor perturbations or (and) of a running in the power spectrum 
of primordial perturbations the best fit of  $\tau$ lies in the range  to $\simeq 0.091-0.092$ (0.096).
While this simple ``$\tau$-parametrization'' of the reionization process and, in particular,
of its imprints on the \acs{CMB} anisotropy likely represents a sufficiently accurate modelling
for the interpretation of current \acs{CMB} data, a great attention has been recently posed on the 
accurate computation of the reionization signatures in the \acs{CMB} for
a large variety of astrophysical scenarios and physical processes
(see e.g. \cite{psh,doroshkevich02,cen2003,ciardi03,doroshkevich03,kasuya,hansen,popa05,wyithe})
also in the view of \acs{WMAP} accumulating data and of forthcoming and future experiments beyond \acs{WMAP} (see \cite{buriganaetal04a} for a review).
In \cite{paperI} a detailed study of the impact of reionization, and the associated 
radiative feedback, on galaxy formation and of the corresponding detectable signatures has been presented, focussing
on a detailed comparison of two well defined alternative prescriptions
({\it suppression} and {\it filtering})
for the radiative feedback mechanism suppressing star formation in small-mass halos, 
showing that 
they are consistent with a wide set of existing observational data but predict 
different $21$ cm background signals accessible to future observations.
The corresponding signatures detectable in the \acs{CMB} have been then computed in \cite{paperII}.\\
\indent Different scenarios have been investigated in 
\cite{pavel} assuming that structure formation and/or extra sources of energy injection in the cosmic  plasma can induce a double reionization epoch of the Universe at low ({\it late} processes) or high ({\it early} processes) redshifts, providing suitable analytical representations, called hereafter as `` phenomenological reionization histories".
In the late models, hydrogen was typically considered firstly ionized at a higher redshift ($z \sim 15$,  
mimicking a possible effect by Pop III stars) and then at lower redshifts ($z \sim 6$,
mimicking the effect by stars in galaxies), while in the early reionization framework the authors hypothesized a {\it peak like reionization} induced by energy injection in the cosmic plasma at $z \gsim {\rm some} \times 10^{2}$.\\
\indent
In this work we present a detailed analysis of these relatively wide sets of astrophysical and phenomenological models and of the signatures they induce in the \acs{CMB} temperature and polarization anisotropies, 
but the methods described here can be used as guidelines for the implementation of any other reionization scenario (see e.g. \cite{salvaterra_etal_11}). 
In particular, we present original computations for the polarization B-mode angular power spectrum (\acs{APS}) and compare them with the
sensitivity  of on-going ({\it Planck}\footnote{www.rssd.esa.int/Planck}, see \cite{blue_book}) and future (assuming \acs{COrE}\footnote{http://www.core-mission.org/}, see \cite{core_whitepap},  as a reference) \acs{CMB} space missions. \\

\subsection{Signatures in the \acs{CMB}}

\indent The cosmological reionization leaves imprints on the \acs{CMB} depending on the
(coupled) ionization and thermal history. 
They can be divided in three categories\footnote{Inhomogeneous reionization 
also produces \acs{CMB} secondary anisotropies that dominate over
the primary \acs{CMB} component for $l\gsim 4000$ and can  be detected by upcoming
experiments, like the \acf{ACT} or \acf{ALMA} 
\cite{salvaterra,iliev}.}: 
$i)$ generation of \acs{CMB} Comptonization and free-free spectral distortions
associated to the \acf{IGM} electron temperature increase during the reionization epoch     
\cite{BE,ZIS72,ZS69,procopioburigana09,fixsen96},
$ii)$ suppression of \acs{CMB} temperature anisotropies at large multipoles, $\ell$,
due to photon diffusion, and $iii)$ increasing
of the power of \acs{CMB} polarization and temperature-polarization
cross-correlation anisotropy at various multipole ranges,
mainly depending on the reionization epoch,
because of the delay of the effective last scattering surface.
The imprints on \acs{CMB} anisotropies are mainly dependent on the ionization history while \acs{CMB} spectral distortions strongly depend also on the thermal history. 
The reionization process mainly influences the polarization E and B modes, and the cross-correlation temperature-polarization mode because of the linear polarization induced by the Thomson scattering.
The effect is typically particular prominent  at low multipoles $\ell$, showing as a bump in the power spectra, otherwise missing.

Through this note we assume a flat $\Lambda$CDM model compatible with \acs{WMAP}, described by
matter and cosmological constant (or dark energy) density parameters $\Omega_m=0.24$ and $\Omega_\Lambda=0.76$, 
reduced Hubble constant $h=H_0/(100{\rm km/s/Mpc)}=0.73$, baryon density $\Omega_b h^2=0.022$, density contrast
$\sigma_8=0.74$, and adiabatic scalar perturbations (without running) 
with spectral index $n_s=0.95$. We adopt a \acs{CMB} background temperature of $2.725$K \cite{mather99}.

\section{Astrophysical reionization models}
\label{astromodel}

The 
analysis of Ly$\alpha$ absorption in the spectra of the 19 highest redshift \acf{SDSS} \acf{QSO}
shows a strong evolution of the Gunn-Peterson Ly$\alpha$ opacity at $z \sim 6$ \cite{Fan2006, Gallerani2006}. 
The downward revision of the electron scattering optical depth to $\tau = 0.09 \pm 0.03$ in the release of
the 3yr \acs{WMAP} data, confirmed by subsequent releases,
is consistent with ``minimal reionization models'' which do not require the presence of
very massive ($M>100 M_\odot$) Pop III stars \cite{Gnedin2006}.
The above models can be then used
to explore the effects of reionization on galaxy formation, referred to as ``radiative feedback".

A semi-analytic model to jointly study cosmic
reionization and the thermal history of the \acs{IGM} has been developed in \cite{Choudhury2005}. According to Schneider and collaborators, the semi-analytical model developed
by Choudhury \& Ferrara, complemented by the additional physics introduced in \cite{Choudhury2006}, 
involves: $i)$ the treatment of \acs{IGM} inhomogeneities by adopting
the procedure of \cite{Miralda2000}; $ii)$ the 
modelling of the \acs{IGM} treated as a multiphase medium, following the thermal
and ionization histories of neutral, HII, and HeIII regions simultaneously
in the presence of ionizing photon sources represented by Pop III stars with a standard Salpeter \acs{IMF}
extending in the range $1-100~M_\odot$ \cite{Schneider2006}, Pop II stars with $Z=0.2 Z_{\odot}$
and Salpeter \acs{IMF}, and \acsp{QSO} (particularly relevant at $z \lsim 6$);
$iii)$ the chemical feedback controlling the prolonged transition from Pop III to Pop II stars
in the merger-tree model by Schneider;  
$iv)$ assumptions on the escape fractions of ionizing photons, considered to be independent of
the galaxy mass and redshift, but scaled to the amount of produced ionizing photons.
It then accounts for radiative feedback inhibiting star formation in low-mass galaxies. 
This semi-analytical model is determined by only four free parameters: 
the star formation efficiencies of Pop II
and Pop III stars, a parameter, $\eta_\mathrm{esc}$, related to the escape fraction of ionizing
photons emitted by Pop II and Pop III stars, and the normalization of the photon
mean free path, $\lambda_0$, set to reproduce low-redshift observations of Lyman-limit 
systems.

A variety of feedback mechanisms can suppress star formation in mini-halos, 
i.e. halos with virial temperatures $T_{vir} < 10^4$~K, particularly if their 
clustering is taken into account \cite{Kramer2006}. It is then possible to assume that stars 
can form in halos down to a virial temperature of $10^4$~K, consistent with the 
interpretation of \acs{WMAP} 
data 
\cite{Haiman2006} (but see also \cite{Alvarez2006}). 
Even galaxies with virial temperature $T_{vir} \gsim 10^4$~K can be significantly 
affected by radiative feedback during the reionization process, as the increase
in temperature of the cosmic gas can dramatically suppress their formation.

Based on cosmological simulations
of reionization, \cite{Gnedin2000} developed an accurate characterization of the radiative 
feedback on low-mass
galaxies. This study shows that the effect of photoionization is
controlled by a single mass scale in both the linear and non-linear regimes.
The gas fraction within dark matter halos  at any given moment is fully specified
by the current filtering mass, which directly corresponds to the length
scale over which baryonic perturbations are smoothed in linear theory. The
results of this study provide a quantitative description of radiative feedback,
independently of whether this is physically associated to photoevaporative flows
or due to accretion suppression.

Two specific alternative prescriptions
for the radiative feedback by these halos have been considered:

\noindent
$i)$ {\it suppression model} -- 
in photoionized regions halos  can form stars
{\it only} if their circular velocity exceeds the critical value
$v_\mathrm{crit} = [{2 k_B T}/{\mu m_\mathrm{p}}]^{{1/2}}$; 
here $\mu$ is the mean molecular weight, $m_\mathrm{p}$ is the proton mass, 
and $T$ is the average temperature of
ionized regions, computed self-consistently from the multiphase \acs{IGM} model;

\noindent
$ii)$ {\it filtering model} -- 
the average baryonic mass $M_{b}$ within halos in photoionized regions
is a fraction of the universal value $f_{b} = \Omega_b/\Omega_m$, given by the fitting formula
${M_b}/{M} = {f_b}/{[1+(2^{1/3}-1) M_\mathrm{C}/M]^3}$,
where $M$ is the total halo mass and $M_\mathrm{C}$ is the total mass of halos  
that on average retain 50\% of their gas mass.
A good approximation for 
$M_\mathrm{C}$ is given by the linear-theory filtering mass,
$M_\mathrm{F}^{2/3} = ({3}/{a}) \int_0^a da^\prime M_\mathrm{J}^{2/3}(a^\prime)
\left[1-\left({a^\prime}/{a}\right)^{1/2}\right]$,
where $a$ is the cosmic scale factor,
$M_\mathrm{J} \equiv ({4 \pi}/{3}) \bar{\rho} \left({\pi c_\mathrm{s}^2}/{G \bar{\rho}}\right)^{3/2}$
is the Jeans mass, $\bar{\rho}$ is the average total mass density of the Universe, 
and $c_\mathrm{s}$ is the gas sound speed.

The model free parameters are constrained by a wide range of observational data. 
Schneider and collaborators reported the best fit choice of the above four parameters for these two 
models that well accomplish a wide set of astronomical observations, such as the 
redshift evolution of Lyman-limit absorption systems, the Gunn-Peterson and electron 
scattering optical depths, the cosmic star formation history, and number counts of 
high redshift sources in the NICMOS Hubble Ultra Deep Field.

The two feedback
prescriptions have a noticeable impact on the overall reionization history and the relative contribution
of different ionizing sources. In fact, although the two models predict similar global star formation
histories dominated by Pop II stars, the Pop III star formation rates have
markedly different redshift evolution. Chemical feedback forces Pop III stars to live preferentially
in the smallest, quasi-unpolluted halos  
(virial temperature $\gsim 10^4$~K), which are
those most affected by radiative feedback.
In the suppression model, where star formation is totally suppressed below 
$v_\mathrm{crit}$, Pop III
stars disappear at $z \sim 6$; conversely, in the filtering model, where halos  
suffer a gradual reduction of the available
gas mass, Pop III stars continue to form at $z \lsim 6$, though with a declining rate.
Since the star formation and photoionization rate at these redshifts are observationally well constrained,
the star formation efficiency and escape fraction of Pop III stars need 
to be lower in the filtering model in
order to match the data. 
Therefore reionization starts at $z \lsim 15$ in the filtering model and only 16\% 
of the volume is 
reionized at $z=10$
(while reionization starts at $z \sim 20$ in the suppression model and it is 85\% 
complete by $z=10$).
For $6 < z < 7$, QSOs, Pop II and Pop III give a comparable contribution 
to the total photo-ionization rate in the filtering model, whereas in the 
suppression model reionization at 
$z < 7$ is driven 
primarily by QSOs, with a smaller
contribution from Pop II stars only. 

The predicted free electron fraction and gas temperature evolution 
in the redshift
range $7 < z < 20$ is very different for the two feedback models.
In particular, in the filtering model the gas kinetic temperature is heated above 
the \acs{CMB} value only at $z \lsim 15$. 

The Thomson optical depth, $\tau = \int \chi_e n_e \sigma_T c dt$,
can be directly computed for the assumed $\Lambda$CDM cosmological model
parameters given the ionization histories:
$\tau_{CF06} \simeq 0.1017$ and $\tau_{G00} \simeq 0.0631$ 
for the suppression and the filtering model, respectively.
Note that these values are consistent with the Thomson optical depth range 
derived from \acs{WMAP} $3$yr ($7$yr) data but with $\sim 1 \sigma$ ($\sim 2 \sigma$) 
difference among the two models, leaving a chance of accurately 
probing them with forthcoming \acs{CMB} anisotropy experiments. 

\subsection{Fitting astrophysical histories}
\label{tool}

\begin{figure}[ht]
    \begin{minipage}{0.48\textwidth}
    \centering
	\includegraphics[scale=0.43]{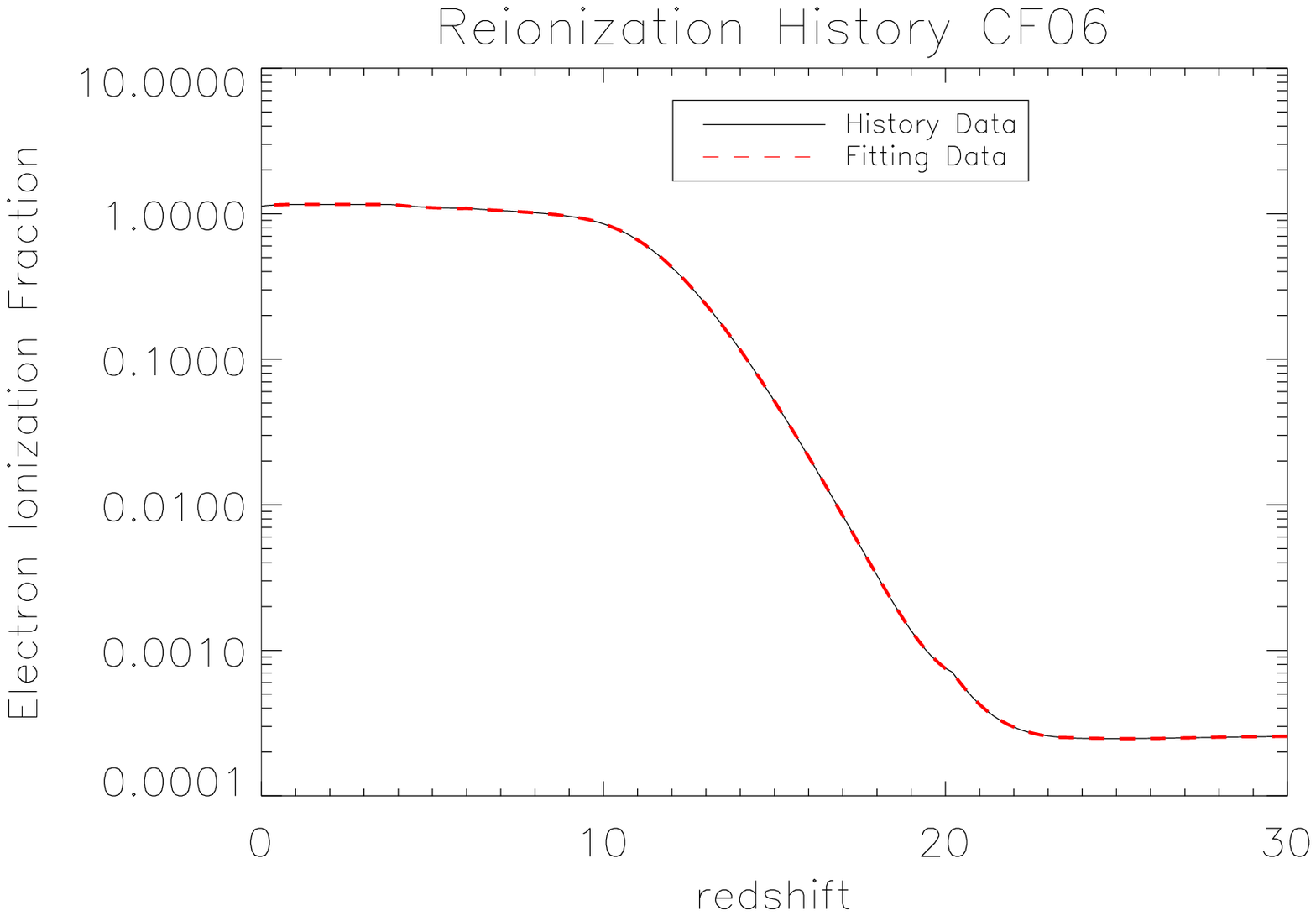}
    \end{minipage}\hfill
    \begin{minipage}{0.48\textwidth}
	\centering
	\includegraphics[scale=0.43]{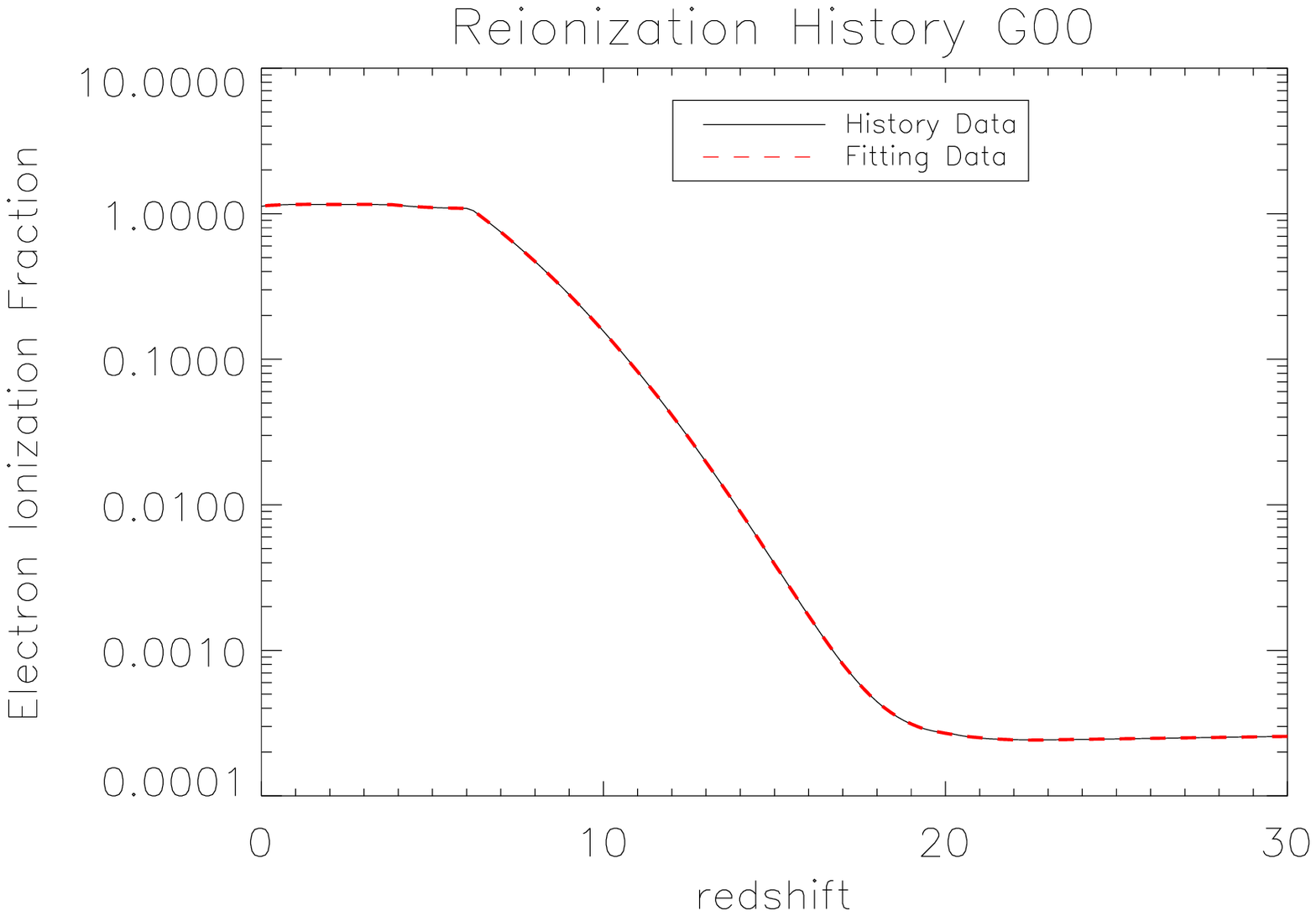}
     \end{minipage}
	\caption{Ionization fraction: comparison between the tabulated data (solid black line) and the fit (dashed red line) for the suppression (left panel) and filtering model (right panel).}
	\label{fig:cfgre}
\end{figure}

\begin{figure}[ht]
    \begin{minipage}{0.48\textwidth}
    \centering
	\includegraphics[scale=0.43]{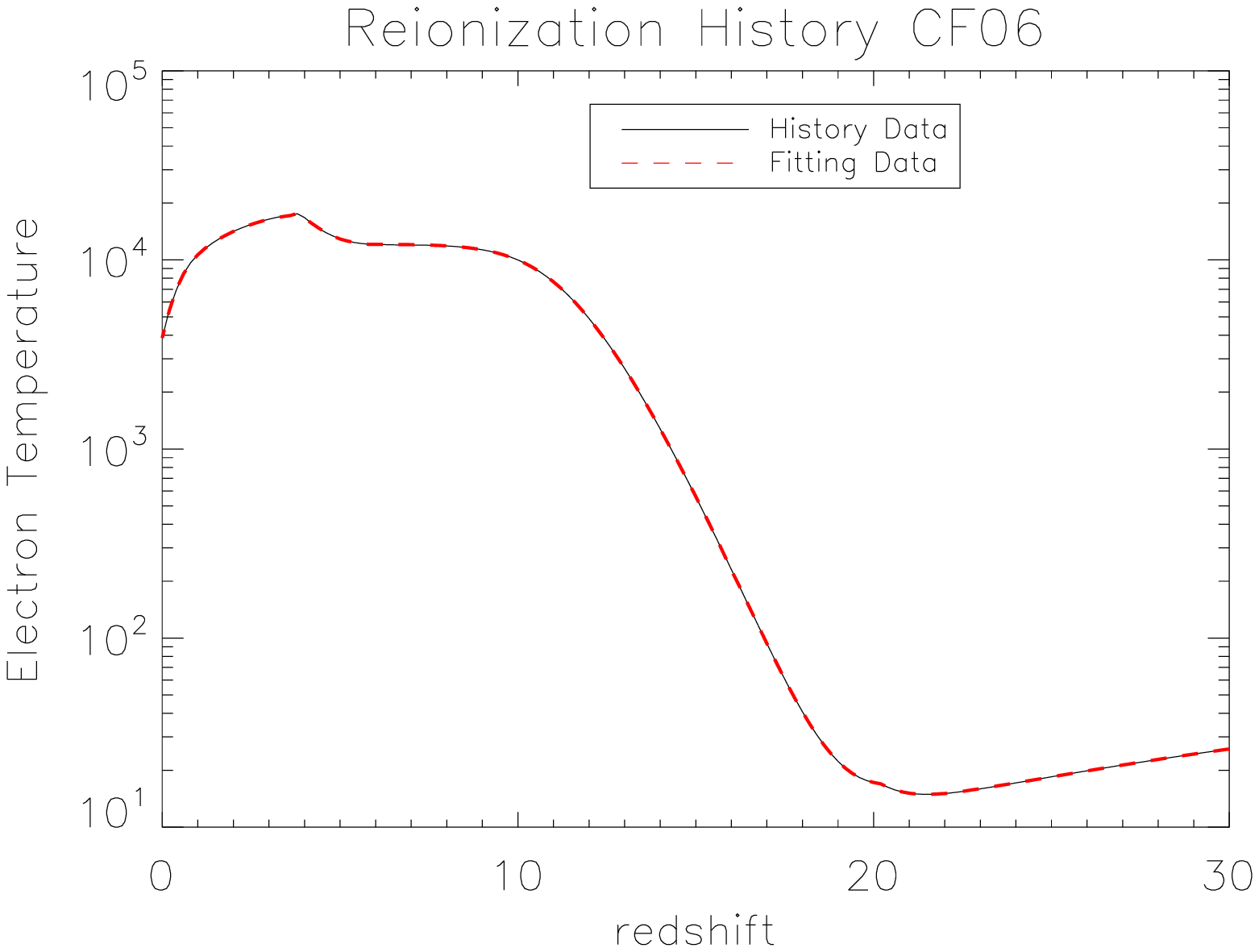}
    \end{minipage}\hfill
    \begin{minipage}{0.48\textwidth}
	\centering
	\includegraphics[scale=0.43]{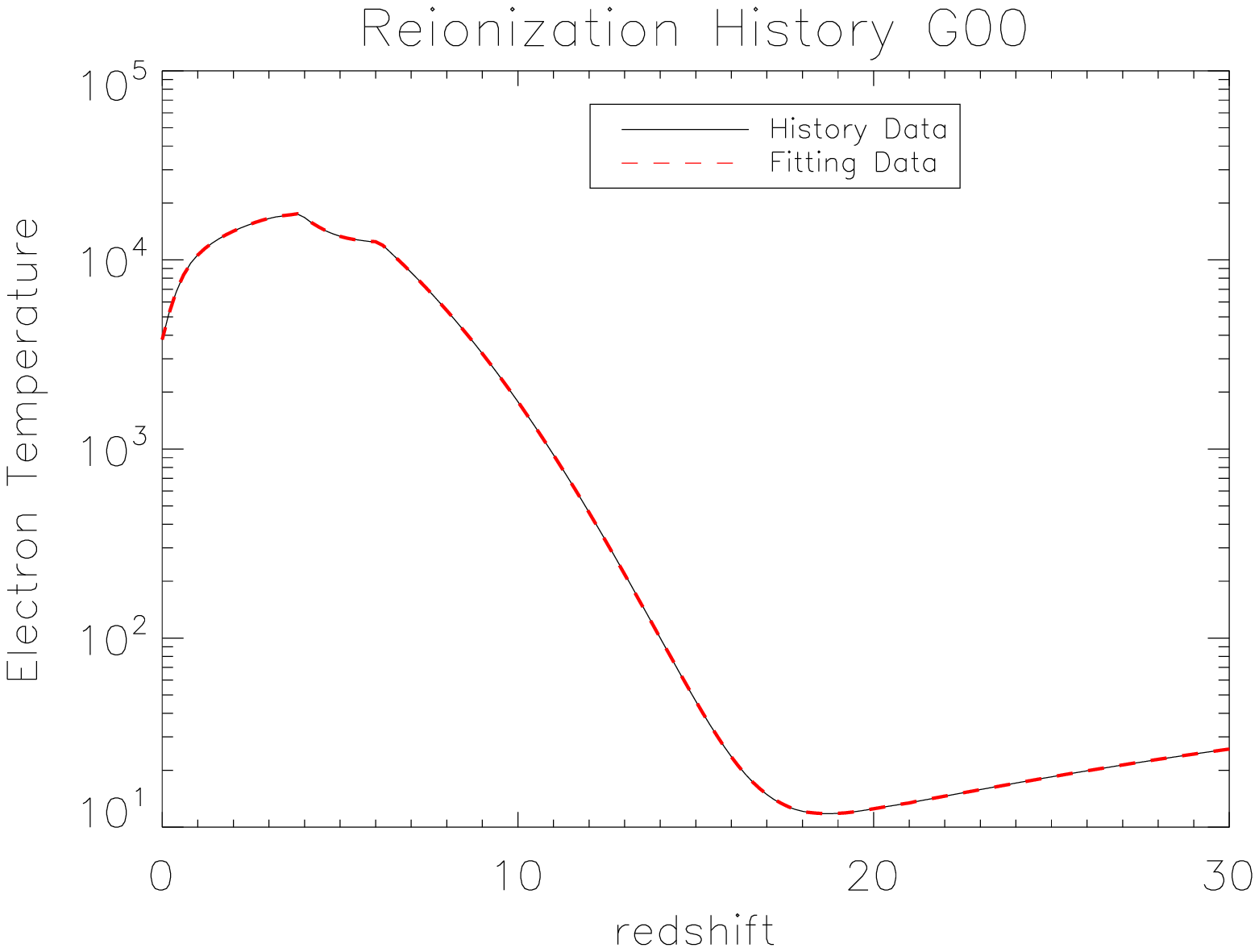}
     \end{minipage}
	\caption{Electron temperature: comparison between the tabulated data (solid black line) and the fit (dashed red line) for the suppression (left panel) and filtering model (right panel).}
	\label{fig:cfgtemp}
\end{figure}

Since filtering and suppression models gives the redshift evolution of ionization fraction and electron temperature in a tabular way, as first step we derived these quantities with appropriate analytical functions, by means of a specific fitting tool,  \emph{Igor Pro (v. 6.21)} \cite{wavemetrics}.
In order to have an accurate parametrization we divided the redshift in bins such to minimize the $\chi ^{2}$ test given by the fit itself (see Appendix \ref{fitting} for details).  \\
The results found using the complete set of functions specific to each epoch of interest are plotted in Fig.~\ref{fig:cfgre} for the ionization fraction and in Fig.~\ref{fig:cfgtemp} for the electron temperature.
Every graph shows the fitting functions (dashed red line) and the corresponding reionization model (solid black line).  

\begin{figure}[here]
    \begin{minipage}{0.48\textwidth}
    \centering
    \includegraphics[scale=0.43]{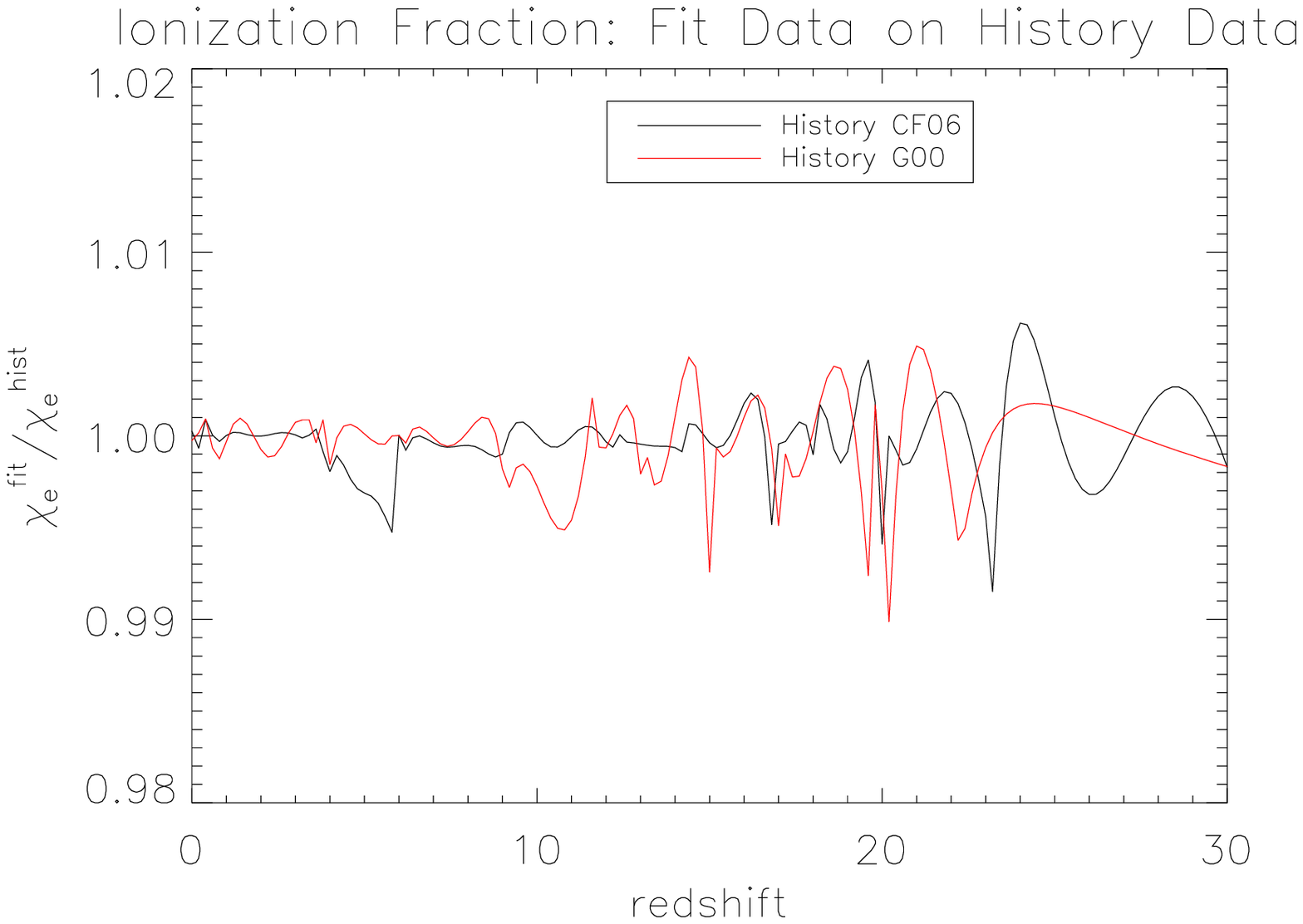}
    \end{minipage}\hfill
    \begin{minipage}{0.48\textwidth}
	\centering
	\includegraphics[scale=0.43]{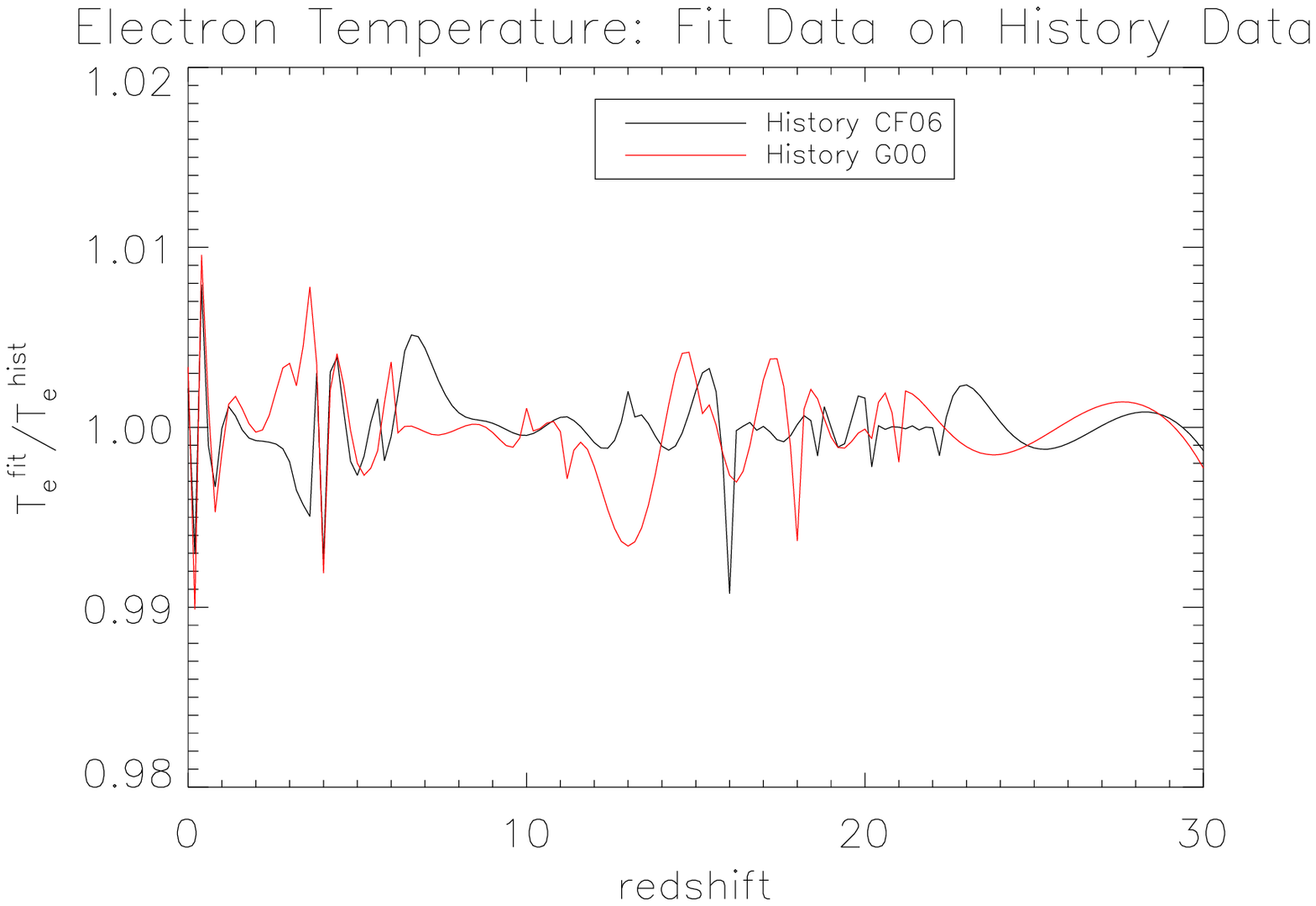}
     \end{minipage}
	\caption{Ionization fraction: ratios between the fit functions and the tabulated data for the ionization fraction (left panel) and electron temperature (right panel) as function of redshift for the two different reionization histories.}
	\label{fig:fracretemp}
     \end{figure}

The accuracy of the fit can be analyzed by means of the percentile difference among theoretical data and fitting functions, namely the ratios between the derived functions and the models data, as shown in Fig.~\ref{fig:fracretemp}. From these plots one can see that the difference is always $ < 1\%$, a bit greater in the filtering model but, in any case, the precision of the fit is always very high. 

\section{Phenomenological reionization models}
\label{phenmodel}

We investigated two phenomenological double peaked reionization models 
introduced by Naselsky \& Chiang
in which the Universe was reionized twice at different epochs, the first one, {\it late reionization model}, at $z \sim 10$ by Pop III stars and then at $z \sim 6$ by stars in large galaxies, and the second one, {\it early reionization model}, at very high redshift ($z \gsim 100$), induced for example by decay of unstable particles, followed by reionization mechanisms  at $z \sim 6$ as predicted in the standard picture. These models are based on the assumption that extra ionizing sources can be described  by the efficiency of the ionizing photons production coefficient $\varepsilon_{i} $ which characterizes  the ionizing photon production rate:\\

\begin{equation}
\frac{dn_{i}}{dt} = \varepsilon_{i}(z) n_{b}(z) H(z),
\end{equation}
\\

\noindent where $n_{b}$ is the baryon density and $H(z)$ is the Hubble parameter.\\
\indent The late reionization history has been modeled as:\\

\begin{equation}
\varepsilon_{i}(z) = \varepsilon_{0} {\rm exp} \left [ -\frac{(z-z_{re})^{2}}{\Delta z^{2}} \right ] + \varepsilon_{1} (1+z)^{-m} \Theta (z_{re}-z).
\label{eq:epslate}
\end{equation} 
\\
\noindent The two terms in the right hand of the equation describe the two ionizing epochs, with $\varepsilon_{0}, \varepsilon_{1}$ and $m$ free parameters of the model,  $\Delta z$ the width of the first reionization epoch, $\Theta$ the step function. The first peak corresponds to a reionization fraction that decreases considerably at $z > z_{re}$, while the second term characterizes a reionization fraction which is monotonic increasing with increasing time (or decreasing redshift). \\
\indent For the early reionization history, the authors adopted a representations in which the efficiency coefficient has a Gaussian parametric form:\\

\begin{equation}
\varepsilon_{i}(z) = \xi {\rm exp} \left [ -\frac{(z-z_{re})^{2}}{\Delta z^{2}} \right ];
\label{eq:epsearly}
\end{equation} 
\\
again, $\xi$, $z_{re}$ and $\Delta z$ are free parameters (clearly different from those of the previous scenario). \\

\begin{figure}[ht]
\centering
\includegraphics[scale=0.5]{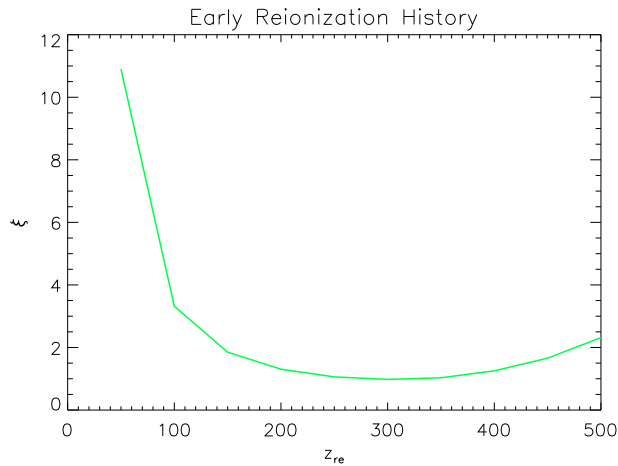}
\caption{Dependence of the free parameter $\xi$ on the reionization redshift of the early history in order to have the adopted value of $\tau$.}
\label{fig:csizre}
\end{figure}

\indent Varying the free parameters of the considered model allows us to identify reionization histories producing a Thomson optical depth with the desired value.  
With this approach we selected for the late history a set of parameters able to approximately reproduce the optical depth found for the suppression model: \\

\begin{equation}
\begin{cases}
\varepsilon_{0} = 1.3 \cdot 10^{3} \\ z_{re} = 10 \\ \beta = \frac{\varepsilon_{1}}{10^{9}} = 5.3 \cdot 10^{-6} \\ m = 11.95 \\  \Delta z = 0.025 z_{re} 
\end{cases}
\quad\Rightarrow\quad \tau_{L} \sim \tau_{CF06} = 0.1017.
\end{equation} 
\\

We exploited also the combination of an early reionization history, unable to contribute to the actual optical depth having a substantial high reionization redshift, with the filtering model such as the global optical depth is exactly the same that figures out from the suppression model: \\

\begin{equation}
\begin{cases}
\xi = 2.315 \\ z_{re} = 500 \\  \Delta z = 0.025 z_{re} 
\end{cases}
\quad\Rightarrow\quad \tau_{E}^{tot} = \tau_{E} + \tau_{G00} = \tau_{CF06} = 0.1017.
\end{equation} 
\\

In Fig.~\ref{fig:csizre} is shown the dependence of the variable $\xi$ on the reionization redshift, $z_{re}$, in order to satisfy the above condition (for the considered choice of  $\Delta z$).

In both astrophysical and phenomenological models we assumed the same cosmological parameters defined in the previous section. \\
\indent The ionization fraction can be evaluated from the balance between the recombination and ionization processes:

\begin{equation}
\frac{d x_{e}}{dt} = - \alpha_{rec}(T) n_{b} x_{e}^{2} + \varepsilon_{i}(z)(1 - x_{e})H(z)
\end{equation} 
\\
\noindent
where $\alpha_{rec} \sim 4 \cdot 10^{-13} (T/10^{4} K)^{-0.6} s^{-1}cm^{-3}$ is the recombination coefficient. \\

\noindent Assuming a curvature term $\Omega_{k} = 0$, and so $\Omega_{\Lambda} = 1 - \Omega_{m}$ the Hubble parameter is approximated  by:

\begin{equation}
H(z) = H_{0} \sqrt{\Omega_{m}(1+z)^{3}+\Omega_{\Lambda}} \, .\\
\end{equation} 

\begin{figure}[ht]
\centering
\includegraphics[scale=0.7]{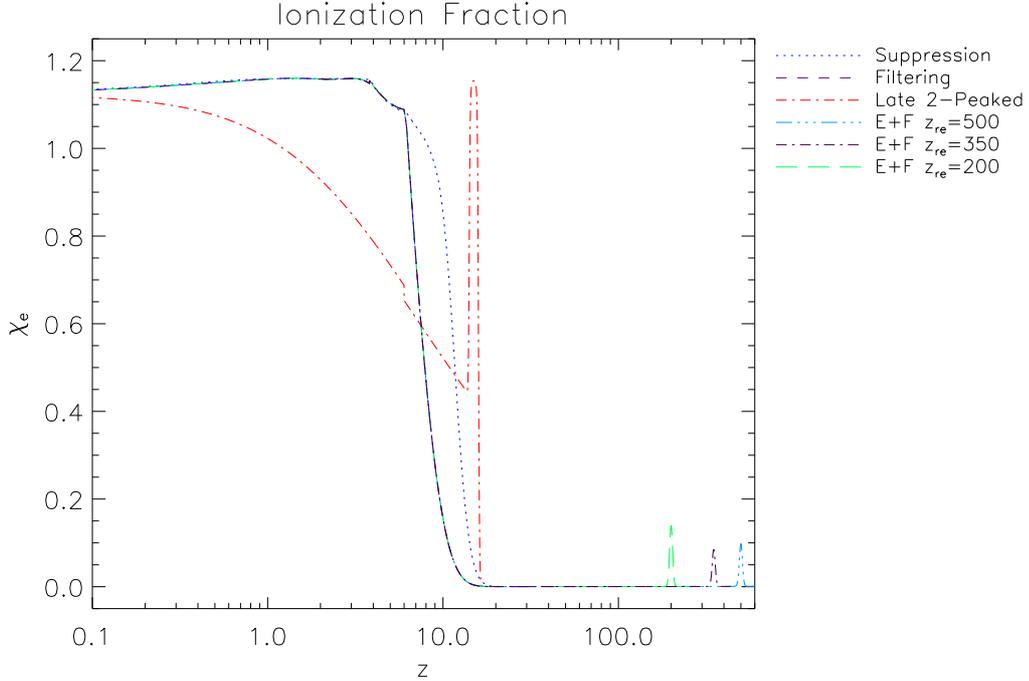}
\caption{Reionization fraction: comparison between models. }
\label{fig:knfrchiE}
\end{figure}

\indent
In Fig.~\ref{fig:knfrchiE} we display the time evolution of the ionization fraction for all the models considered in this work. In particular, the coupled early and filtering model is plotted for  three different cases of the reionization redshift. Note that, in order to have a constant optical depth, we varied the parameter $\xi$ assuming values $\xi = (2.315, 1.031, 1.309)$ when $z_{re} = (500, 350, 200)$ respectively, giving a decreasing in the peak of the high redshift region of $x_{e}$, not linear with the reionization redshift. The reason can be found analyzing the redshift dependence of the recombination and ionization coefficients that are, respectively:
\begin{equation}
C_{rec} \propto n_{b} \propto z^{3},
\end{equation} 
and 

\begin{equation}
C_{ion} \propto \varepsilon_{i}(1 - x_{e})H(z) \propto \xi z^{3/2},
\end{equation} 
so that the higher is $z_{re}$ the lower is $x_{e}$ (see Table~\ref{recion} for details).
\\
\begin{table}[ht]
\centering
\begin{tabular}{| c | c | c | c |}
\hline
$\xi$ & $z_{re}$ & $C_{rec}$ & $C_{ion}$ \\
\hline
$2.315$ & $500$ & $1.25 \cdot 10^{8}$ & $2.59 \cdot 10^{4} $ \\
$1.031$ & $350$ & $4.29 \cdot 10^{7}$ & $6.75 \cdot 10^{3}$ \\
$1.309$ & $200$ & $8.00 \cdot 10^{6}$ & $3.70 \cdot 10^{3} $ \\
\hline
\end{tabular} 
\caption{Redshift dependence of recombination and ionization coefficients for decreasing reionization redshift and adjusted $\xi$ model parameter when joining the early reionization history with the filtering model.}
\label{recion}
\end{table}
\\

In these prescriptions the ionization fraction and electron temperature are provided in an analytical form and connected through the dependence on the temperature of the recombination coefficient $\alpha_{rec}$, that enters in the time evolution equation of the ionization fraction. Since the relatively weak dependence, $\alpha_{rec} \propto T^{-0.6}$, the details of the assumptions on the matter temperature are not particularly critical, although, in principle, for the active phases out of equilibrium, they could play a non negligible role. For the active phase, we assume here, respectively, a temperature profile mimicking the model by Cen in the case of late processes and a temperature profile given by the same Gaussian parametric form as in 
Eq. (\ref{eq:epsearly}) but with a peak temperature, $T_{p}$, as free parameter instead of $\xi$ in the case of early processes, as adopted by Burigana and collaborators in 2004. We use here $T_{p} = 6 \cdot 10^{4}$K. When the ionizing photons production is negligible and no longer affects the ionization history we assume the minimal ionization fraction usually derived in the absence of source terms, thus assuring the continuity with the quiescent phases in the evolution of the plasma properties.

\subsection{Comoving fractions of injected photon number and energy density}

For many mechanisms of cosmological reionization, the underlying physical process is usually characterized in terms of an additional source of ionizing photons injected in the plasma. We  link here the parameters of the considered phenomenological histories to the comoving fractions of injected photon number density and energy density.
Defining the usual \acs{CMB} photon number density $n_{0} \simeq 399 (T_{0} / 2.7$K$)^{3}$, the comoving fractions of photon number density injected in the redshift range $z_{i} \ge z \ge z_{f}$ is given by:\\
\begin{equation}
\frac{\Delta n}{n} = \int_{z_{f}}^{z_{i}} \frac{\varepsilon_{i}(z) n_{b} (z)}{n_{0} (1 + z)^{4}} dz.
\end{equation} 
\\
The second term, $\varepsilon_{1} (1+z)^{-m} \Theta (z_{re}-z)$, appearing in the late reionization history can be easily integrated:
\begin{equation}
\frac{\Delta n}{n} = \int_{0}^{z_{re}} \frac{\varepsilon_{1} n_{b0}}{n_{0} (1 + z)^{1+m}} dz = \frac{\varepsilon_{1} n_{b0}}{n_{0} m} [1 - (1 + z_{re})^{-m}] \, ,
\end{equation} 

\noindent 
where $n_{b0}=1.12 \cdot 10^{-5} \Omega_{b} h^{2}$ is the current baryon number density.\\
For the Gaussian term defining the early model and appearing as first term in the late model, we find the following suitable approximation:\\

\begin{equation}
\frac{\Delta n}{n} = \int_{z_{f}}^{z_{i}} \frac{{\rm const} \cdot n_{b0}}{n_{0}(1 + z)} {\rm exp} \left [ - \frac{(z - z_{re})^{2}}{\Delta z^{2}} \right ] dz \\
\simeq \frac{{\rm const} \cdot n_{b0}}{n_{0}} \frac{\Delta z}{1 + z_{re}} \sqrt{\pi},
\end{equation} 
\\
\noindent where ``const'' is the early parameter $\xi$ or the late parameter $\varepsilon_{0}$, respectively, and the assumption $z_{f} \to -\infty$ and $z_{i} \to \infty$ for the integration limits is made, 
a good approximation to this aim for a peaked Gaussian shape.

The comoving fraction of injected photon energy density depends on the energy distribution function of ionizing photons.
Given the comoving fraction of injected photon number density and assuming the mean ionizing photon energy, $\bar{E}_{\gamma}^{ion}$, here assumed $\simeq 10$ eV for numerical estimates, 
one can compute of the comoving fraction of injected photon energy density, $\Delta \varepsilon/\varepsilon$.
In the case of the early history and for the Gaussian term of the late history we have: \\

\begin{equation}
\frac{\Delta \varepsilon}{\varepsilon} \simeq 1.59 \cdot 10^{4} \frac{\Delta n}{n}  \frac{\bar{E}_{\gamma}^{ion}}{10{\rm eV}}   \left (\frac{T_{0}}{2.7}\right )^{-1}(1+z)^{-1} \, ;
\end{equation} 
\\

\noindent where for numerical estimates we can assume $(1+z) \simeq (1+z_{re})_{early/late}$. For the second term appearing in the late reionization history, after a simple integration we have:\\

\begin{equation}
\frac{\Delta \varepsilon}{\varepsilon} \simeq 1.60 \cdot 10^{-11} \frac{\bar{E}_{\gamma}^{ion}}{10{\rm eV}} \frac{\varepsilon_{1} n_{b0}}{\varepsilon_{0} (m+1)} [1 - (1 + z_{re})^{-(m+1)}] \, ,
\end{equation} 
 
\noindent where $\varepsilon_{0} = a T_{0}^{4}$ is the current \acs{CMB} photon energy density.

In general, we used the $D01AJF$ routine of the \emph{NAG Libraries} for an accurate numerical cross-check of previous analytical formulas and approximations. For the adopted parameters 
we found an agreement better than $\simeq 0.03$\%.

\section{Code implementation}
\label{code}

We modified \acs{CAMB}
\cite{camb}, the cosmological Boltzmann code 
(see also \cite{seliak_zalda_96}) for computing the angular power spectrum  
of the anisotropies of the \acs{CMB}, in order to introduce the 
ionization fractions evaluated according to the astrophysical and phenomenological reionization models described in previous sections,
alternative to the reionization treatment originally implemented in \acs{CAMB}. Of course, the methods described here can be 
used as guidelines for the implementation of any other 
astrophysical reionization model (see Appendix \ref{codecamb} for further details).\\
\indent As a  significant step forward with respect to previous analyses, 
the emphasis of this work is posed to the extension to a first detailed 
characterization of the polarization B-mode \acs{APS}.

By implementing the source file {\it $reionization.f90$}, that defines the {\it Reionization module}, we are able to parametrize the desired reionization history and to supply the corresponding ionization fraction as function of redshift \cite{trombetti_burigana_rep589}.

Particular care must be taken to the normalization of the quantities occurring in the history definition, such as the ionization fraction.
In \acs{CAMB} the reionization fraction is referred to the hydrogen, so, when allowing for helium reionization, the global ionization fraction in the case of complete ionization can be greater than one, following the relation:\\

\begin{equation}
\chi_{e}^{full} = 
1 + \frac{n_{He}}{n_{H}}\, . 
\end{equation}

Instead, in the two astrophysical reionization models, the global ionization fraction corresponding to the case of complete ionization 
is $\chi_{e}^{full,CF06} = 1.12721$ for the suppression model, and $\chi_{e}^{full,G00} = 1.12480$ for the filtering model. \\
\indent To evaluate the total fraction in the case of the phenomenological histories we have normalized it with \acs{CAMB} implementation, such as:\\

\begin{equation}
x_{e}^{CAMB} = \frac{n_{e}}{n_{H}} = x_{e}^{L,E} \frac{1+f_{H}}{2f_{H}}    \, , 
\end{equation}
\\
\noindent
where $f_{H}$ is the fraction of baryonic mass in hydrogen.
Assuming hydrogen abundance $f_{H} = 0.76$, the global ionization fraction in the case of complete ionization for both histories is $f_{L,E} = 1.157895$.\\
\indent Furthermore, it  is possible to fix in \acs{CAMB} the Thomson optical depth parameter, and let the code estimates the reionization redshift, or to choose the desired reionization redshift, and obtaining from dedicated function the $\tau$ evaluation. While for the considered astrophysical models $\tau$ is known, we implemented a modified version of this function to derive $\tau$ according to the considered phenomenological model.  

\subsection{Astrophysical models and standard \acs{CAMB}}

In the framework of our adapted version of \acs{CAMB} code it can be useful to analyze an important differences between the various models we described. 
The crucial characteristic resides in the interplay between different physical processes at various epochs which contribute to reionize the cosmological plasma. Every mixture of these processes can lead to a distinctive scenario, as displayed in Fig.~\ref{fig:knfrchiE}.

\begin{figure}[ht]
\begin{minipage}{0.48\textwidth}
\centering
\includegraphics[scale=0.43]{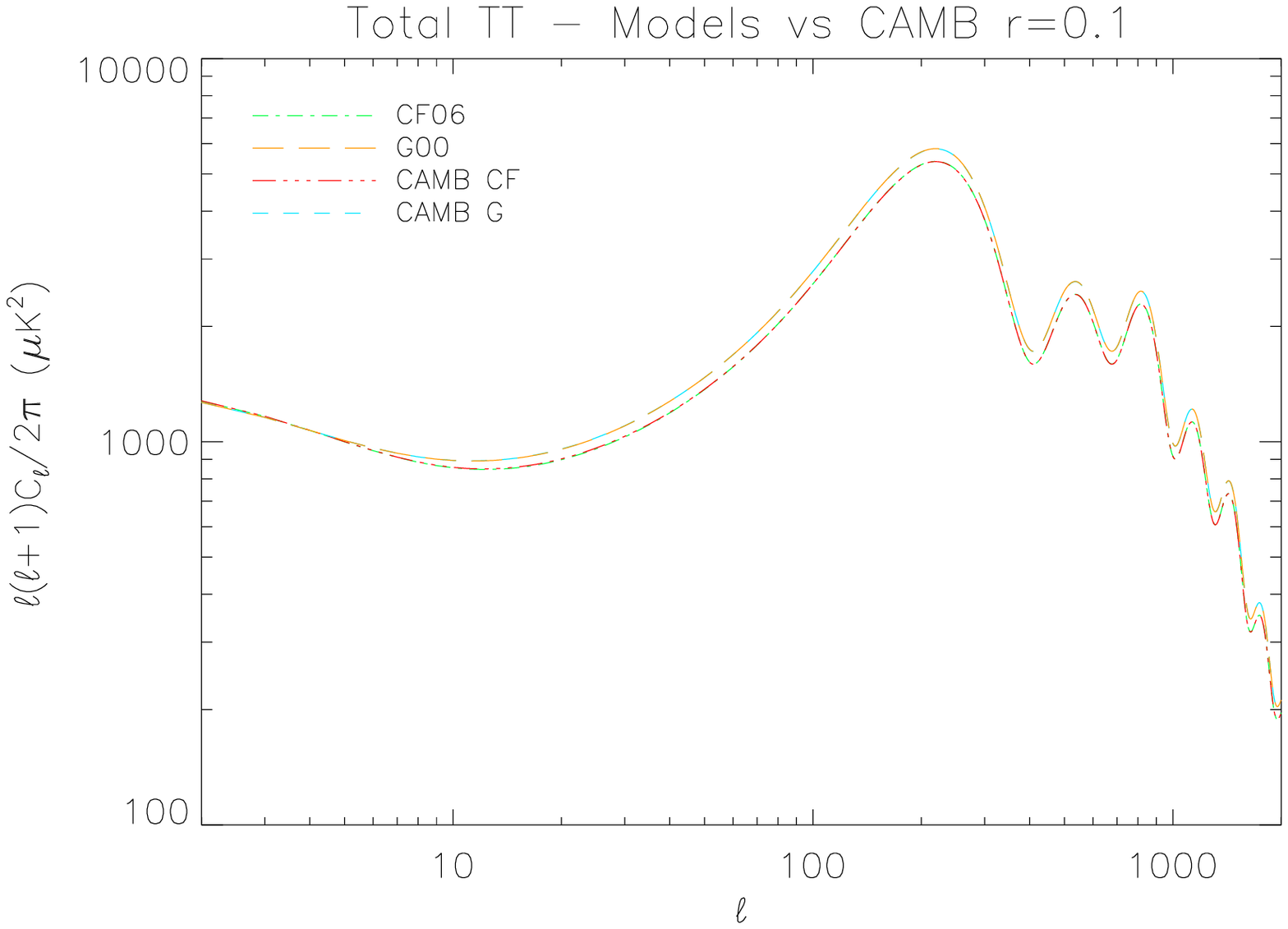}
\end{minipage}\hfill
\begin{minipage}{0.48\textwidth}
\centering
\includegraphics[scale=0.43]{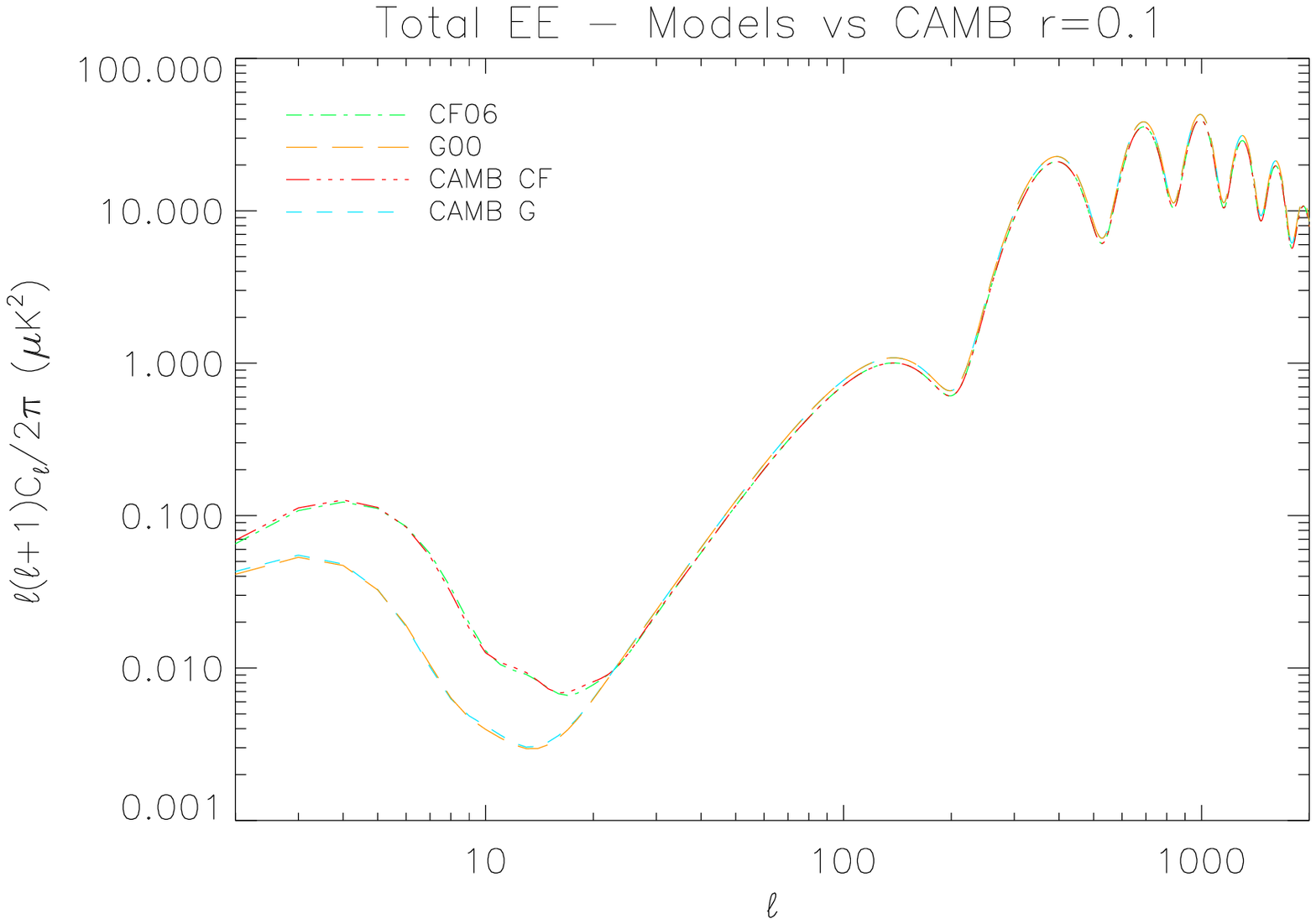}
\end{minipage}
\caption{Comparison between the models and \acs{CAMB} in Temperature (left panel) and E-mode polarization (right panel) \acs{APS}.}
\label{fig:ttee}
\end{figure}

\begin{figure}[ht]
\begin{minipage}{0.48\textwidth}
\centering
\includegraphics[scale=0.43]{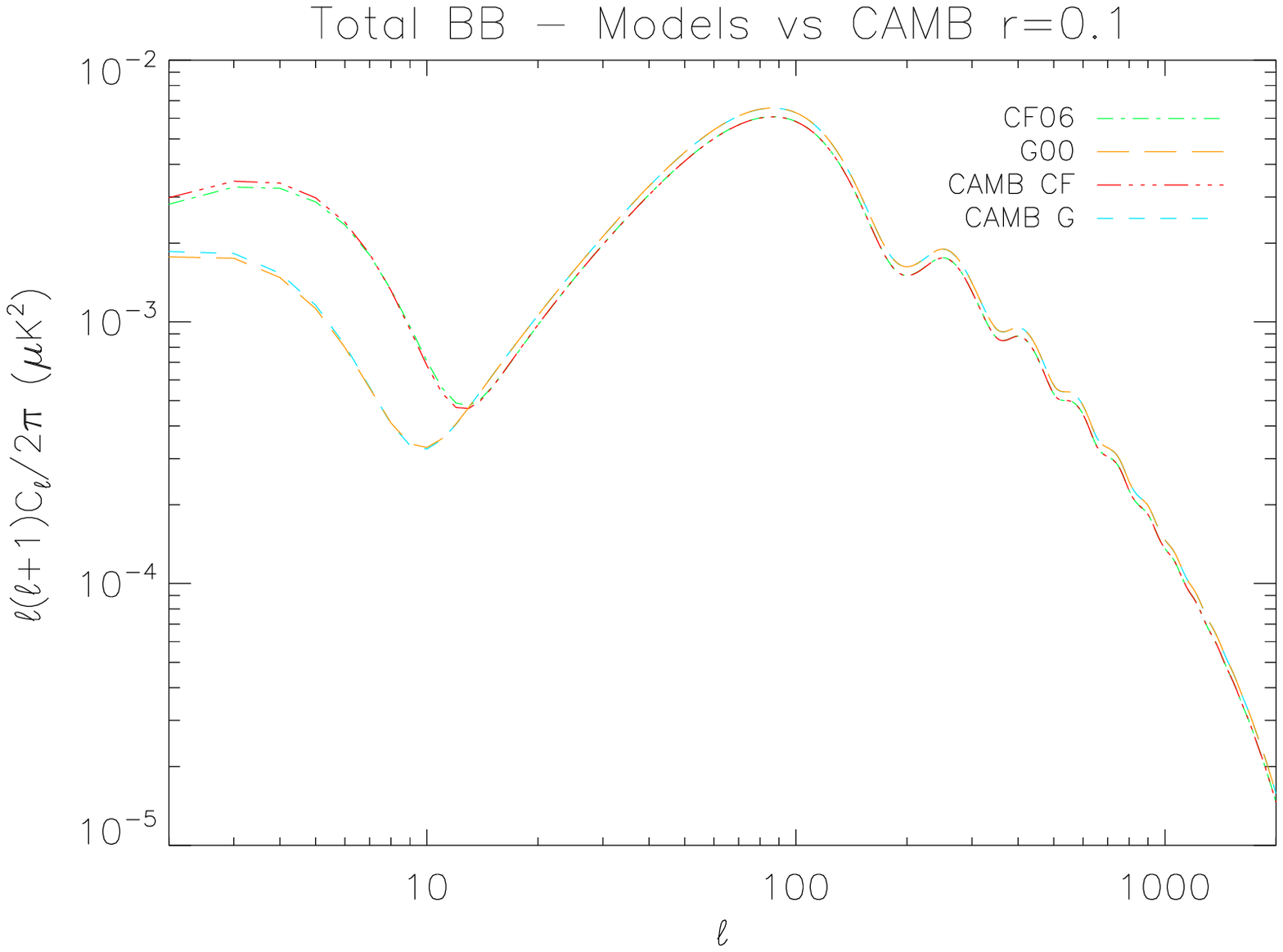}
\end{minipage}\hfill
\begin{minipage}{0.48\textwidth}
\centering
\includegraphics[scale=0.43]{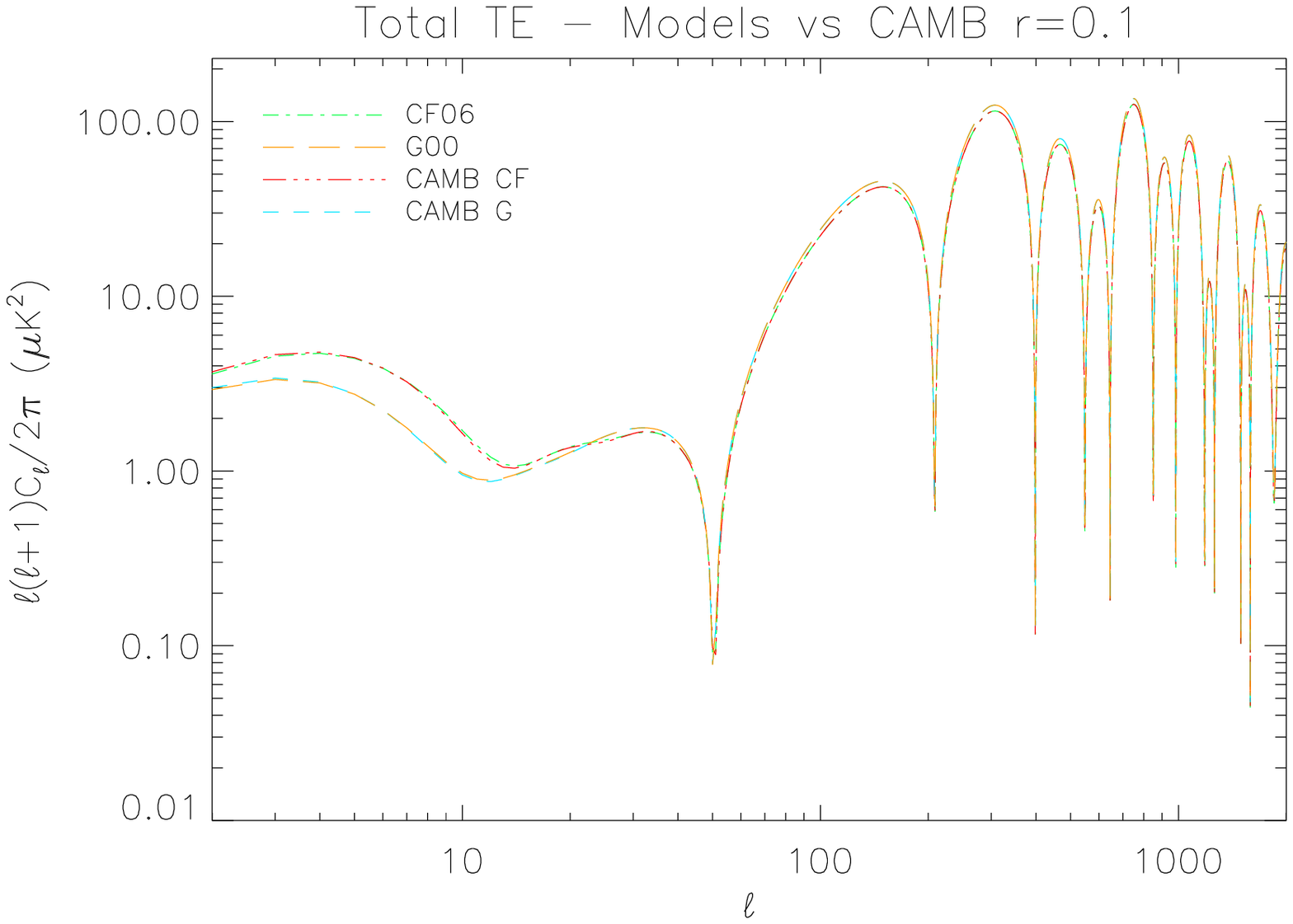}
\end{minipage}
\caption{Comparison between the models and \acs{CAMB} in B-mode polarization (left panel) and temperature-polarization cross-correlation (right panel) \acs{APS}.}
\label{fig:bbte}
\end{figure}

Since the standard \acs{CAMB} traces a reionization that follows the ``single peak'' evolution scheme, it can be interesting to compare it with the suppression and filtering models, investigating on their differences 
in the \acs{APS} and, in particular, in the sequence of acoustic peaks, at various multipoles (see Figs.~\ref{fig:ttee} and~\ref{fig:bbte}). \\
\indent To this aim we derived the temperature (TT), polarization (EE and BB), and cross-correlation (TE) \acs{APS} of \acs{CMB} anisotropies for the two astrophysical reionization histories, \acf{CF06} and \acf{G00}, generically denoted as ``models'' in titles and the legends, and for the original version of the code, denoted as ``\acs{CAMB} CF'' or ``\acs{CAMB} G'',  assuming, respectively, an optical depth corresponding to the value given by the theoretical model to which we are comparing to. Note also that in the right panel of Fig. \ref{fig:bbte} we display the module of the cross-correlation \acs{APS}.

For simplicity, we neglect lensing in this case, and the total $C_\ell$ plotted represents the sum of scalar and tensor contributions. 
The tensor to scalar ratio of primordial perturbation, $r$, is assumed here equal to $0.1$. 

\begin{figure}[ht]
\begin{minipage}{0.48\textwidth}
\centering
\includegraphics[scale=0.43]{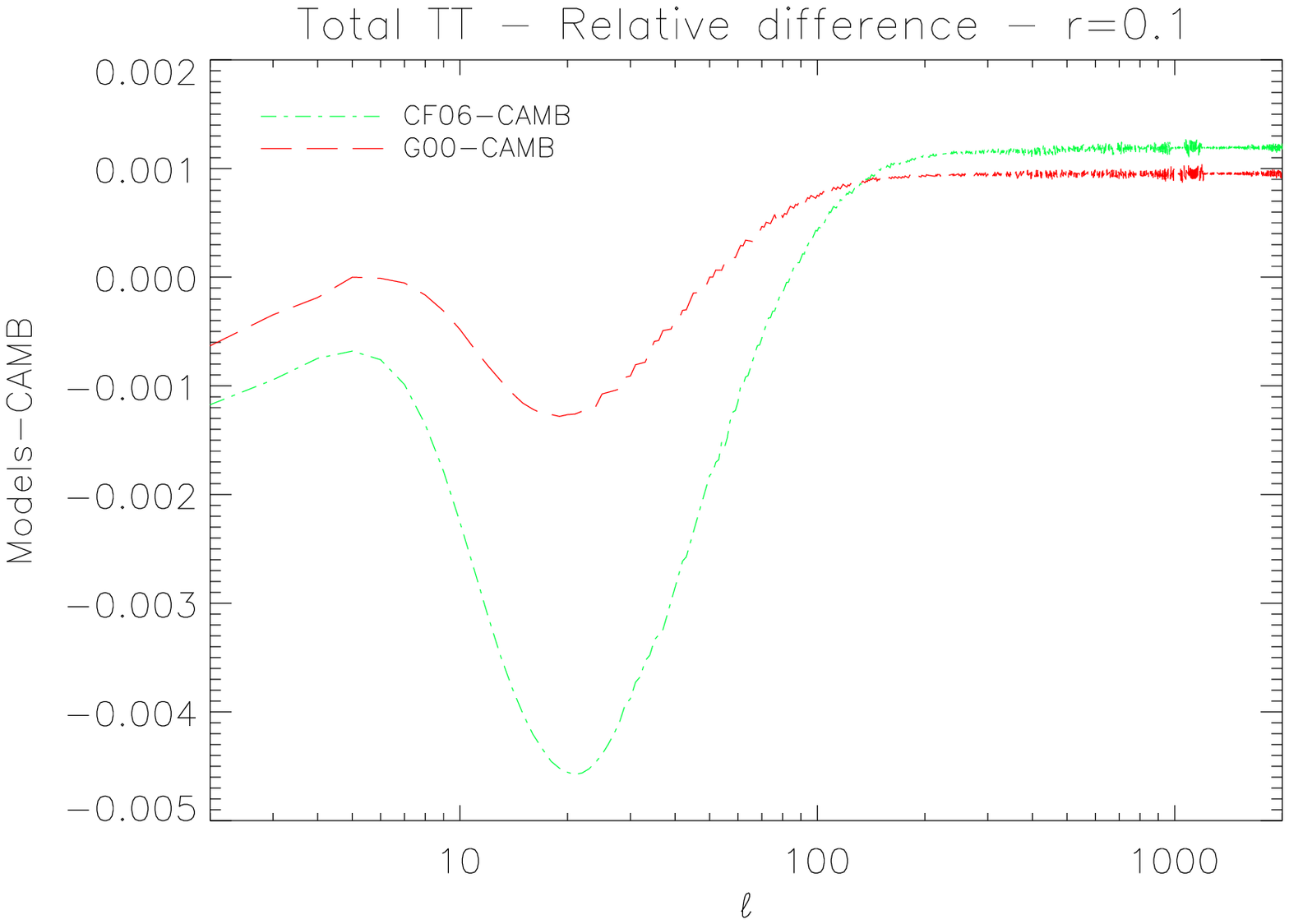}
\end{minipage}\hfill
\begin{minipage}{0.48\textwidth}
\centering
\includegraphics[scale=0.43]{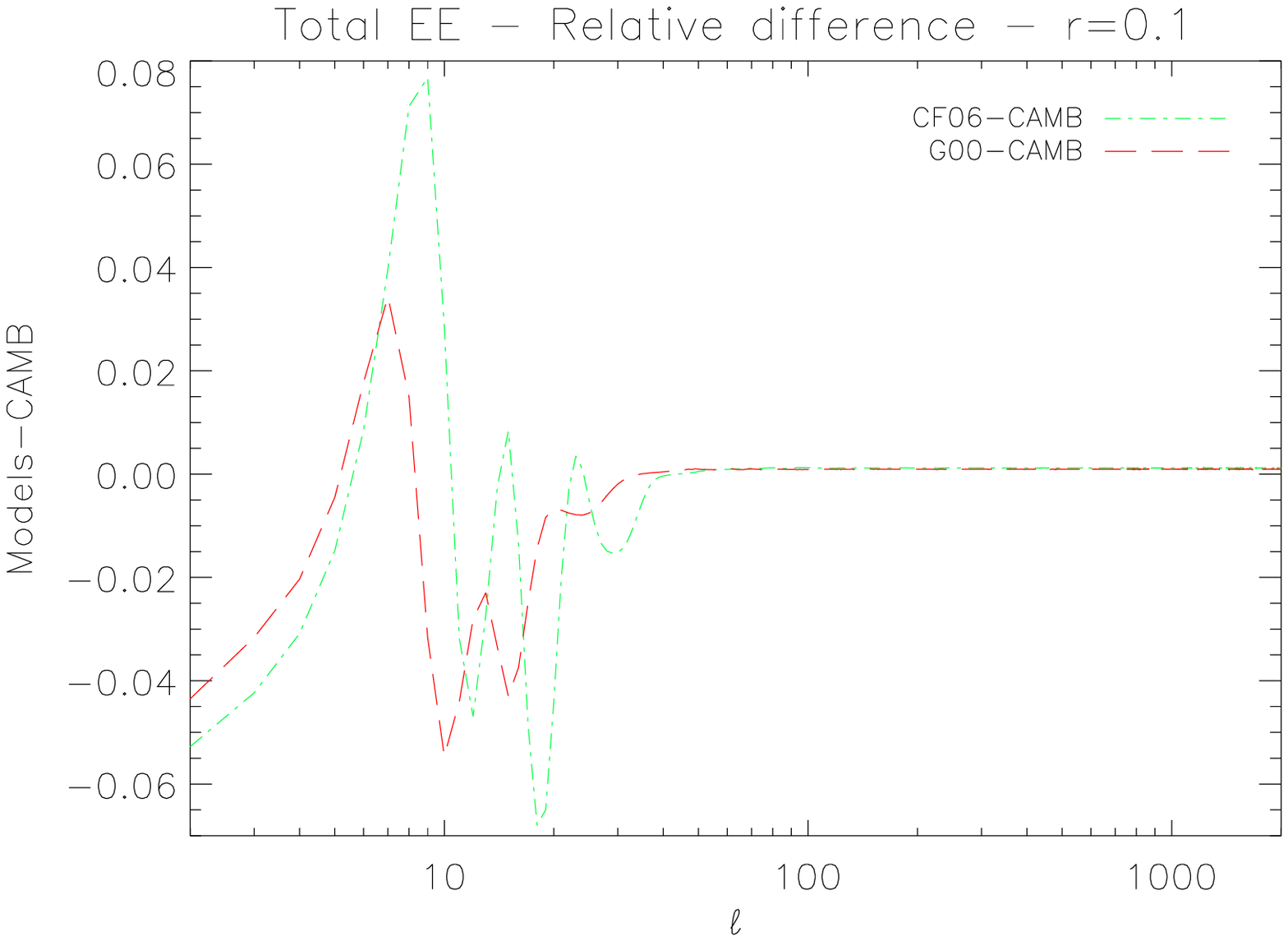}
\end{minipage}
\caption{TT (left panel) and EE (right panel) \acs{APS}: relative differences between the models and \acs{CAMB}.}
\label{fig:reldifttee}
\end{figure}

\begin{figure}[ht]
\begin{minipage}{0.48\textwidth}
\centering
\includegraphics[scale=0.43]{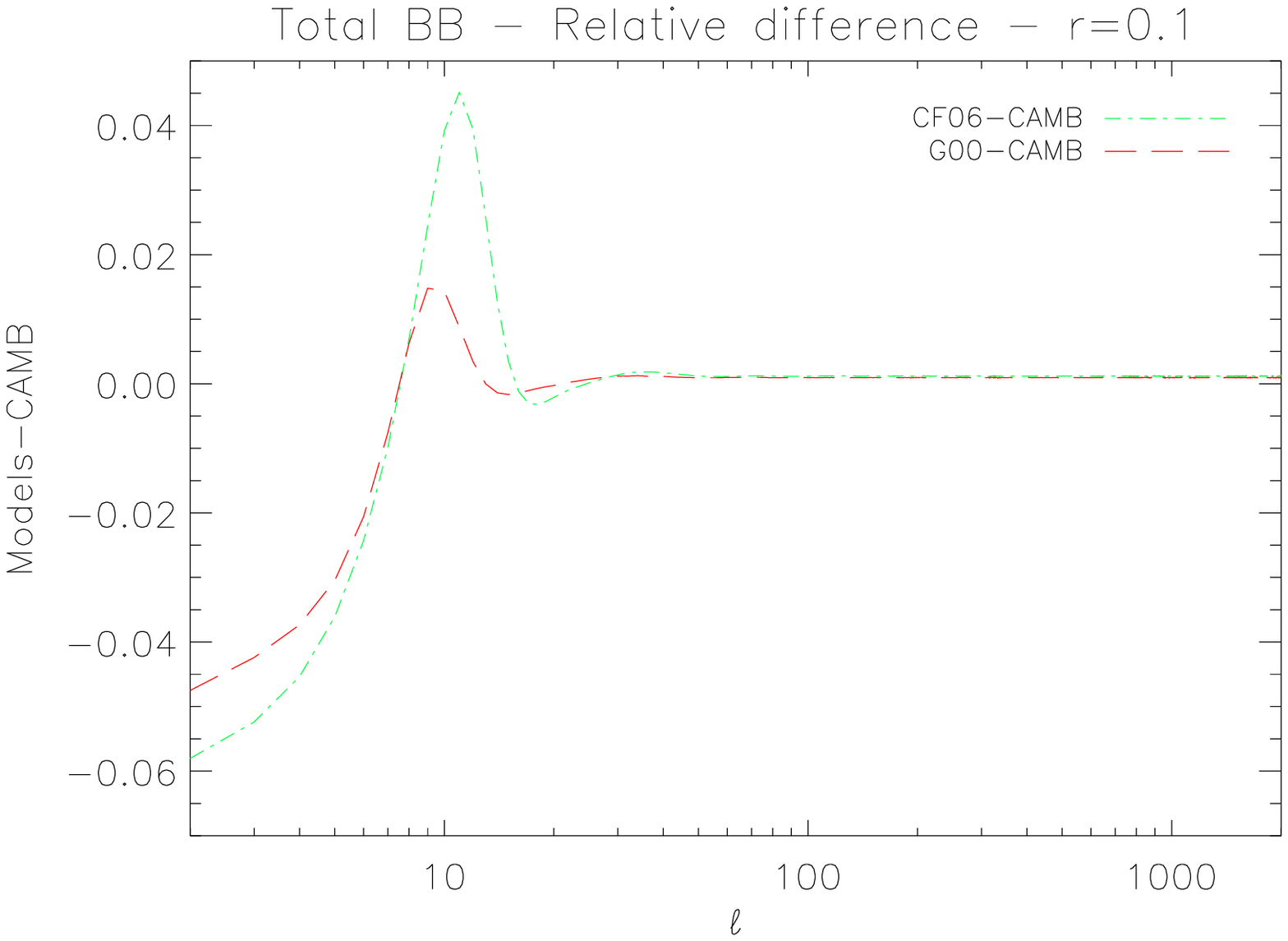}
\end{minipage}\hfill
\begin{minipage}{0.48\textwidth}
\centering
\includegraphics[scale=0.43]{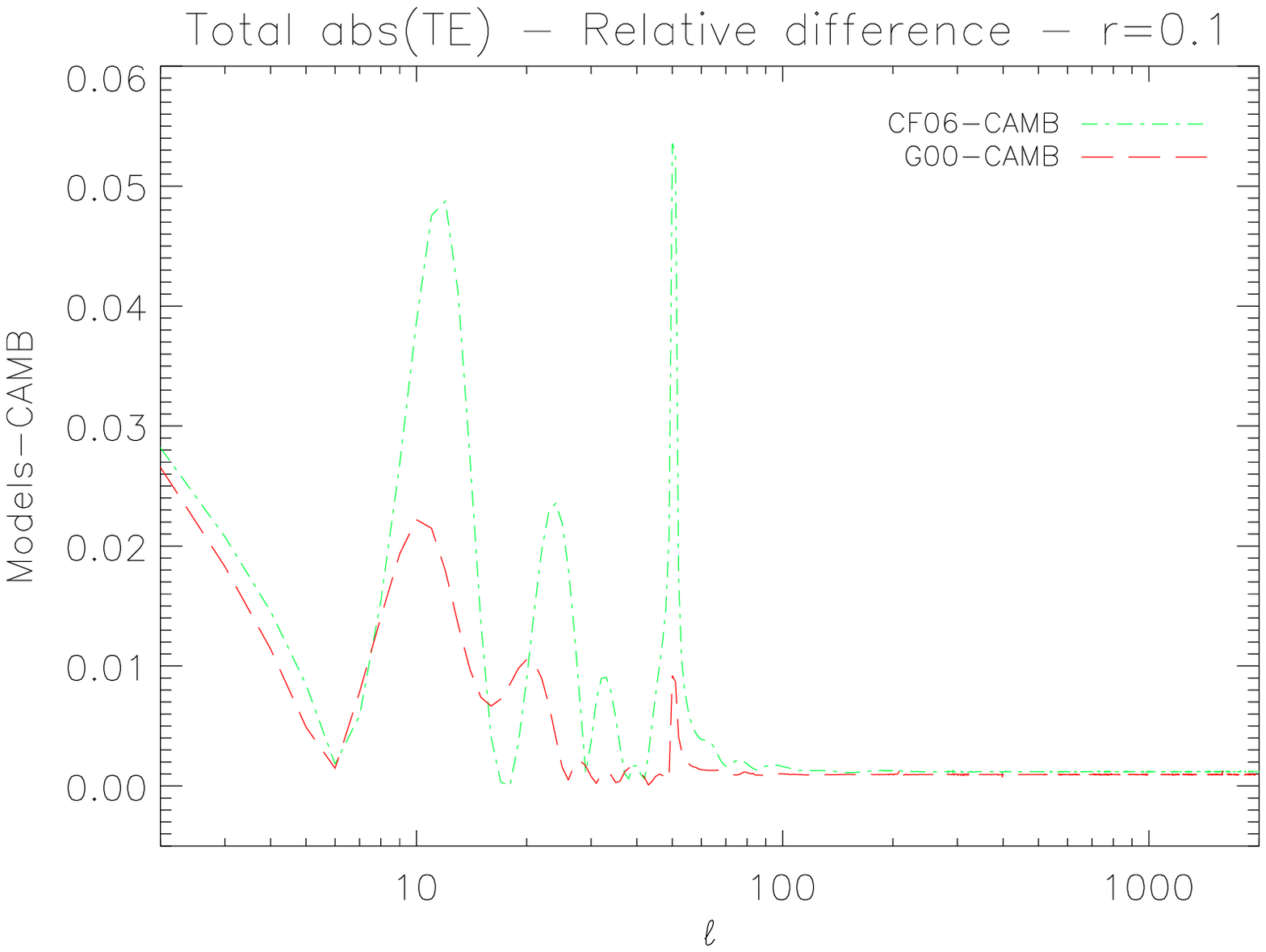}
\end{minipage}
\caption{BB (left panel) and TE (right panel) \acs{APS}: relative differences between the models and \acs{CAMB}.}
\label{fig:reldifbbte}
\end{figure}

It is also very useful to analyse the relative difference between results obtained with the astrophysical models and the original \acs{CAMB}, as shown in Figs.~\ref{fig:reldifttee} e ~\ref{fig:reldifbbte}, defined by the relation:\\

\begin{equation}
\frac{C_{\ell}^{Model} - C_{\ell}^{CAMB}}{1/2(C_{\ell}^{Model} + C_{\ell}^{CAMB})}. \, 
\end{equation} 
\\ 

As before, note that in Fig. \ref{fig:reldifbbte} we display the module of the difference of the cross-correlation \acs{APS}.

The relative differences found in the case of the suppression model are larger than in the case of the filtering prescription, 
mainly in the polarization and cross-correlation patterns of the \acs{APS}. 
In addition, the difference is remarkable at low multipoles, in particular at $\ell< $ few tens, i.e. at intermediate and large angular scales, 
as intuitively expected since we are considering 
relatively late reionization processes.

\section{Sensitivity of \acs{CMB} measurements and foreground emission}
\label{limitations}

\acs{CMB} experimental data are affected by uncertainties due to instrumental noise (crucial at high multipoles, $\ell$, i.e. small angular scales), cosmic and sampling variance (crucial at low $\ell$, i.e. large angular scales) and from systematic effects.
Also, they are contaminated by a significant level of foreground emission of both Galactic and extragalactic origin. In polarization, the most critical Galactic foregrounds are certainly synchrotron and thermal dust emission,
while free-free emission is negligible in polarization and other components, like spinning dust and haze, are still poorly known, particularly in polarization. 
Synchrotron emission is the dominant Galactic foreground signal at low frequencies, up to $\sim 60$GHz, where dust emission starts to dominate. 
External galaxies are critical only at high $\ell$, and radiogalaxies are likely the most crucial in polarization up frequencies $\sim 200$GHz, most suitable for \acs{CMB} anisotropy experiments.

In this section we provide simple recipes aimed at evaluating the levels of sensitivity of on-going and future \acs{CMB} space experiments and the kind of contamination expected from foregrounds, and to make us able discuss in the next section on the possibility to distinguish between different reionization scenarios in the framework of current and future experiments, focussing in particular on the pure polarization E and B-modes.

\subsection{Sensitivity measurements}

The uncertainty on the angular power spectrum is given by the combination of three components, \acf{CV}, \acf{SV} and Instrumental Noise (N) \cite{knox}:\\

\begin{equation}
\frac{\delta C_{\ell}}{C_{\ell}} = \sqrt{\frac{2}{f_{sky}(2 \ell + 1)}}  \left ( 1+ \frac{A \sigma^{2}}{N C_{\ell} W_{\ell}}  \right ).
\end{equation} 

\noindent
Here $f_{sky}$ is the sky coverage, $A$ is the surveyed area, $\sigma$ is the instrumental rms noise per pixel, $N$ is the total pixel number, $W_{\ell}$ is the beam window function that, in the case
of a Gaussian symmetric beam, is:\\

\begin{equation}
W_{\ell} = \rm{exp} (-\ell (\ell + 1) \sigma_{B}^{2}),
\end{equation} 

\noindent being $\sigma_{B} = FWHM/\sqrt{8 \ln 2}$ the beamwidth which defines the experiment angular resolution.

For $f_{sky} = 1$ the first term in parenthesis defines the cosmic variance, an intrinsic limit on the accuracy at which the angular power spectrum of a certain cosmological model defined by a suitable set of parameters
can be measured with the \acs{CMB}. It typically dominates the uncertainty on the \acs{APS} at low $\ell$ because of the small, $2\ell + 1$,
number of modes $m$ for each $\ell$. 
The second term in parenthesis characterizes the instrumental noise, that never vanishes in the case of real experiments. Note also the coupling between experiment sensitivity and resolution, the former defining the 
low $\ell$ experimental uncertainty, namely for $W_{\ell}$ close to unit, the latter determining the exponential loss in sensitivity at angular scales comparable with the beamwidth.\\
\indent In order to provide concrete estimates of these quantities, we consider {\it Planck} \acs{LFI} and \acs{HFI}  channels at $\nu =$ ($70$, $100$, $143$, $217$) GHz, and \acs{COrE} channels at $\nu =$ ($75$, $105$, $135$, $165$, $195$, $225$) GHz, i.e. at the frequencies particularly suitable for \acs{CMB} analyses because of the combination of good experimental sensitivity and resolution, and of relatively low foreground 
contamination. We adopt here the sensitivities and resolutions summarized in the \acs{COrE}  white paper\footnote{The nominal sensitivity of {\it Planck} is slightly better than that adopted here thanks the slightly longer extension of the mission with both instruments, and the additional extension with the only \acs{LFI}. Of course, the real sensitivity of the whole mission will have to include also the potential residuals of systematic effects.}.\\

\begin{table}[ht]
\centering
\begin{tabular}{| c | c | c | c |}
\hline
$\nu$ (GHz) & \acs{FWHM} (arcmin) & $\sigma_{T}$ ($\mu$K $\cdot$ arcmin) & $\sigma_{Pol}$ ($\mu$K $\cdot$ arcmin)  \\
\hline
$70$ & $13$ & $211.2$ & $298.7$ \\
$100$ & $9.9$ & $31.3$ & $44.2$ \\
$143$ & $7.2$ & $20.1$ & $33.3$ \\
$217$ & $4.9$ & $28.5$ & $49.4$ \\
\hline
\end{tabular} 
\caption{Instrumental sensitivity of {\it Planck} experiment.}
\label{sensP}
\end{table}

\begin{table}[ht]
\centering
\begin{tabular}{| c | c | c | c |}
\hline
$\nu$ (GHz) & \acs{FWHM} (arcmin) & $\sigma_{T}$ ($\mu$K $\cdot$ arcmin) & $\sigma_{Pol}$ ($\mu$K $\cdot$ arcmin)  \\
\hline
$75$ & $14$ & $2.73$ & $4.72$ \\
$105$ & $10$ & $2.68$ & $4.63$ \\
$135$ & $7.8$ & $2.63$ & $4.55$ \\
$165$ & $6.4$ & $2.67$ & $4.61$ \\
$195$ & $5.4$ & $2.63$ & $4.54$ \\
$225$ & $4.7$ & $2.64$ & $4.57$ \\
\hline
\end{tabular} 
\caption{Instrumental sensitivity of \acs{COrE} experiment.}
\label{sensC}
\end{table}
\indent For each of the two projects we computed an overall sensitivity value, weighted over the channels, defined by

\begin{equation}
\frac{1}{\sigma_{j,tot}^{2}} = \sum_{i} \frac{1}{\sigma_{j,i}^{2}}.
\end{equation} 

\noindent where $j = T$, $Pol$ and $i$ states for the sensitivity of each frequency channel, listed in Tables~\ref{sensP} and~\ref{sensC}.
\acs{FWHM} values of $13'$ and $14'$ are used to define the overall resolution respectively of {\it Planck} and \acs{COrE} in the computation of the beam window 
function\footnote{In fact, it is possible to smooth maps acquired 
at higher frequencies with smaller beamwidths to the resolution corresponding to the lowest frequency of each experiment.}. \\

\indent Finally, to improve the signal to noise ratio in the \acs{APS} sensitivity, especially at high multipoles, we will apply a multipole binning of $5\%$ in temperature \acs{APS}, $15\%$ in TE cross-correlation and $30\%$ in polarization \acs{APS}, both in E and B-modes.\\

\subsection{Parametrization of residual polarized foreground contamination}

The parametrization of the \acs{APS} of Galactic thermal dust and synchrotron emission adopted in this work is taken from the results of \acs{WMAP} $3$-yrs~\cite{page} under the assumption 
that these contributions are uncorrelated\footnote{A more sophisticated treatment of the foreground power spectra takes into account the correlation among the various foreground components.},
and is expressed by:\\

\begin{equation}
\frac{\ell(\ell + 1)}{2\pi} C_{\ell}^{fore} = (\mathcal{B}_{s}(\nu/65)^{2\beta_{s}} + \mathcal{B}_{d}(\nu/65)^{2\beta_{d}}) \ell^{m}, 
\end{equation} 

\noindent
where $s$ and $d$ stands for synchrotron and dust, and the frequency $\nu$ is expressed in GHz.
The coefficients characterizing the E and B-modes polarization \acs{APS} are slightly different, and are listed in Table~\ref{sincdustEB}.

\begin{table}[ht]
\centering
\begin{tabular}{| c | c | c | c | c | c | c | c |} 
\hline
E-mode & $\mathcal{B}$ ($\mu$K)$^{2}$ & $\beta$ & $m$ & B-mode & $\mathcal{B}$ ($\mu$K)$^{2}$ & $\beta$ & $m$ \\
\hline
sync & $0.36$ & $-3.0$ & $-0.6$ & sync & $0.3$ & $-2.8$ & $-0.6$ \\
dust & $1.00$  & $1.5$  & $-0.6$  & dust  & $0.5$ & $1.5$ & $-0.6$ \\
\hline
\end{tabular} 
\caption{Parametrization of  E and B mode polarization power spectra of Galactic synchrotron and thermal dust emission.}
\label{sincdustEB}
\end{table}

In the next sections we will parametrize a potential residual from non perfect cleaning of \acs{CMB} maps from Galactic foregrounds simply assuming that a certain fraction of the foreground signal {\it at map level} 
(or, equivalently, its square at power spectrum level) contaminates \acs{CMB} maps. Of course, one can easily rescale the following results to any fraction of residual foreground contamination. 
The frequency of 70 GHz, i.e. the {\it Planck} channel where Galactic foreground is expected to be minimum at least at angular scales above $\sim $ one degree, will be adopted as reference. 

For what concerns extragalactic source fluctuations \cite{toffol98,dez05}, we will adopt the recent (conservative) estimate of their Poissonian contribution to the \acs{APS} \cite{tucci_toffol_2012}
at 100 GHz\footnote{We adopt here a frequency slightly larger than that considered for Galactic foregrounds because at small angular scales, where point sources are more critical, the minimum of foreground contamination is likely shifted at higher frequencies.} assuming a detection threshold of $\simeq 0.1$ Jy, together with a potential residual coming from an uncertainty in the subtraction of this contribution computed assuming a relative uncertainty of $\simeq 10$\% in the knowledge of their degree of polarization and in the determination of the source detection threshold, 
implying a reduction to $\simeq 30$\% of the original level \cite{debe_bpol}. Except at very high multipoles, their residual is likely significantly below that coming from Galactic foregrounds.

\section{Results}
\label{results}

In this section we present the angular power spectra resulting from different prescriptions of the reionization process and 
compare them with the sensitivities of on-going and future space experiments including instrumental noise and fundamental statistical uncertainty. \\

\begin{figure}[ht]
\centering
\includegraphics[scale=0.9]{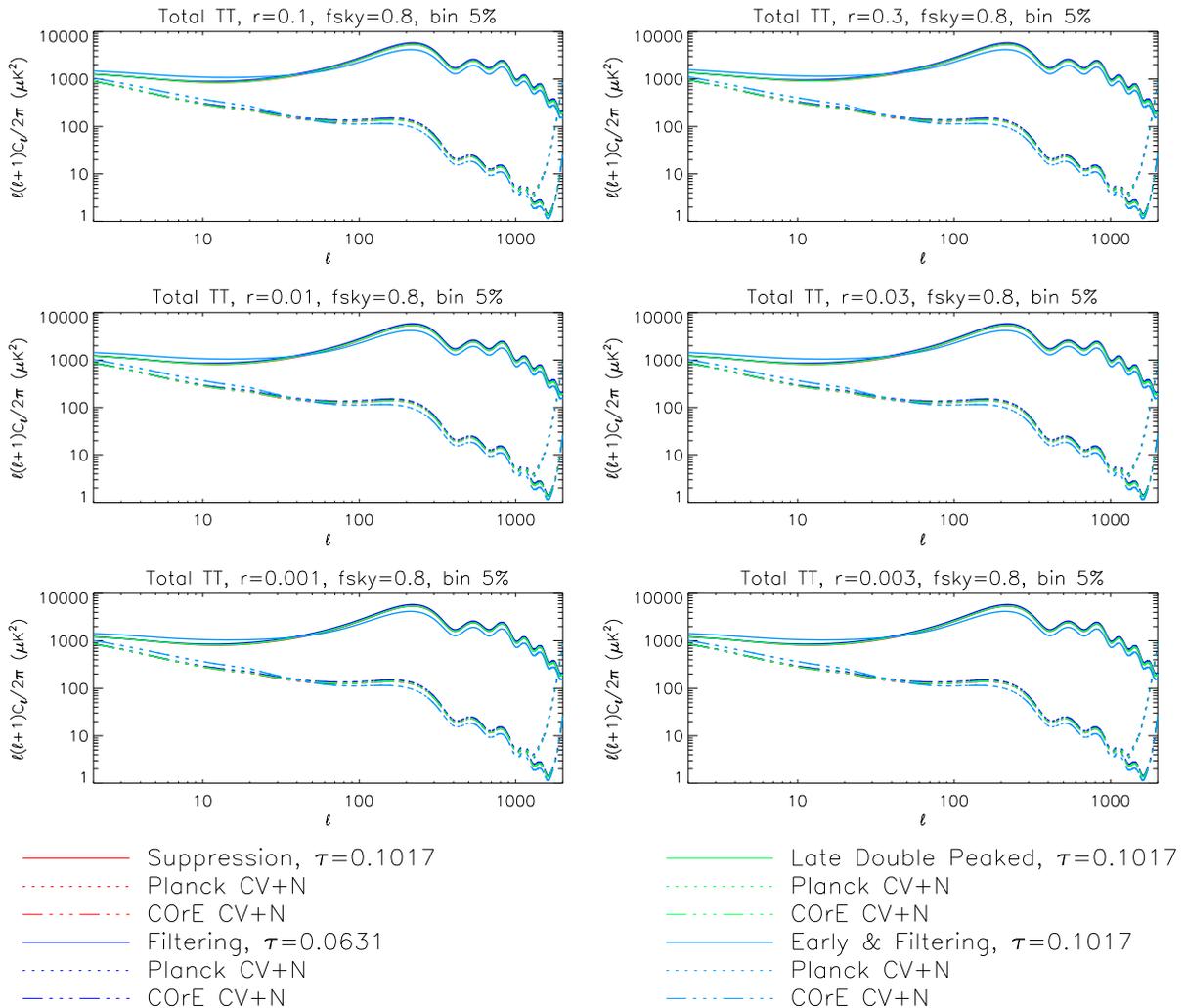}
\vspace*{4.5\baselineskip}
\caption{TT \acs{APS} for the reionization histories: suppression, filtering, late double peaked, early plus filtering for different values of tensor to scalar ratio (see plot legend and text for details).}
\label{fig:ttaps}
\end{figure}

\begin{figure}[ht]
\centering
\includegraphics[scale=0.9]{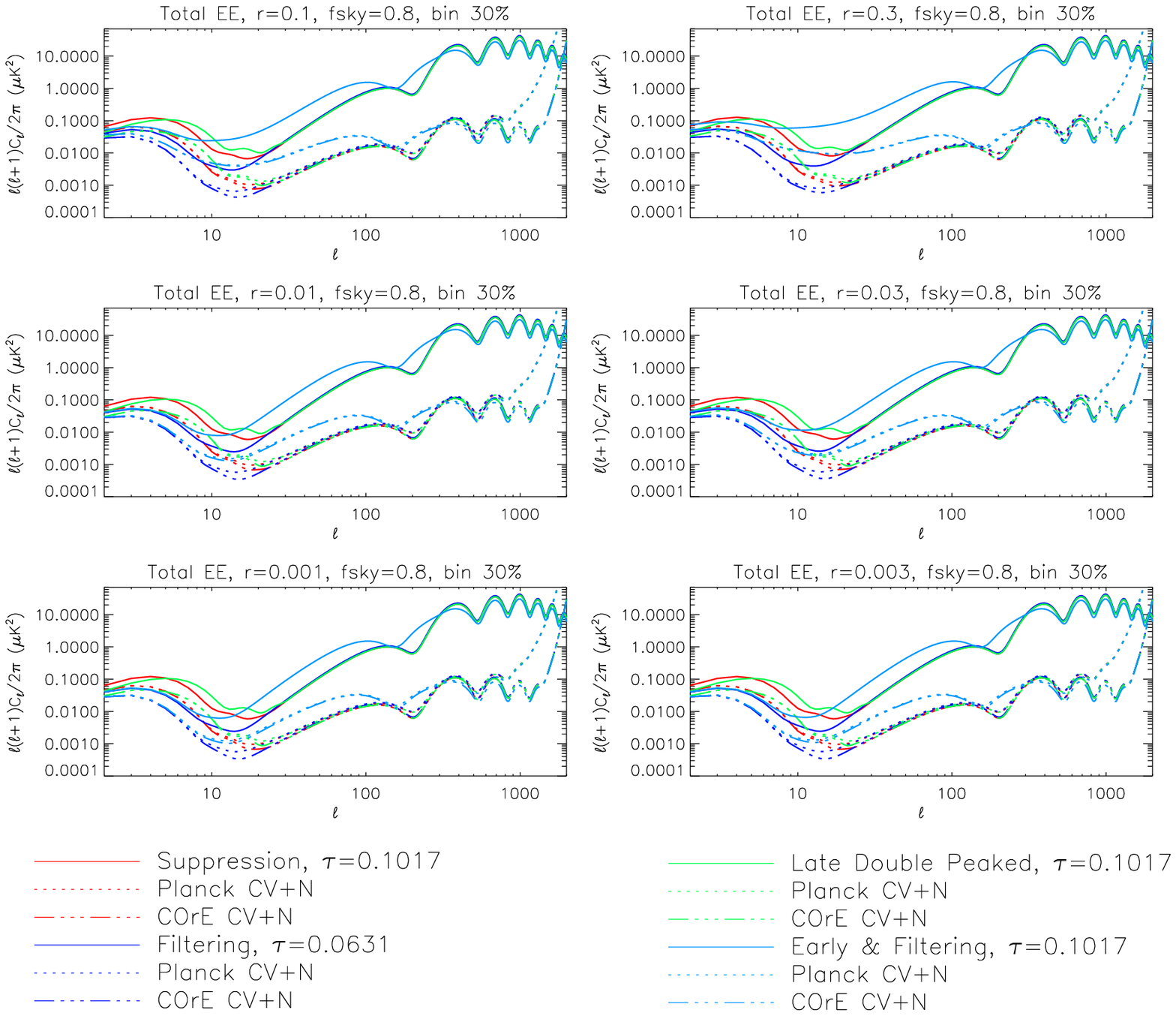}
\vspace*{4.5\baselineskip}
\caption{EE \acs{APS} for the reionization histories: suppression, filtering, late double peaked, early plus filtering for different values of tensor to scalar ratio $r$ (see plot legend and text for details).}
\label{fig:eestory}
\end{figure}

\begin{figure}[ht]
\centering
\includegraphics[scale=0.9]{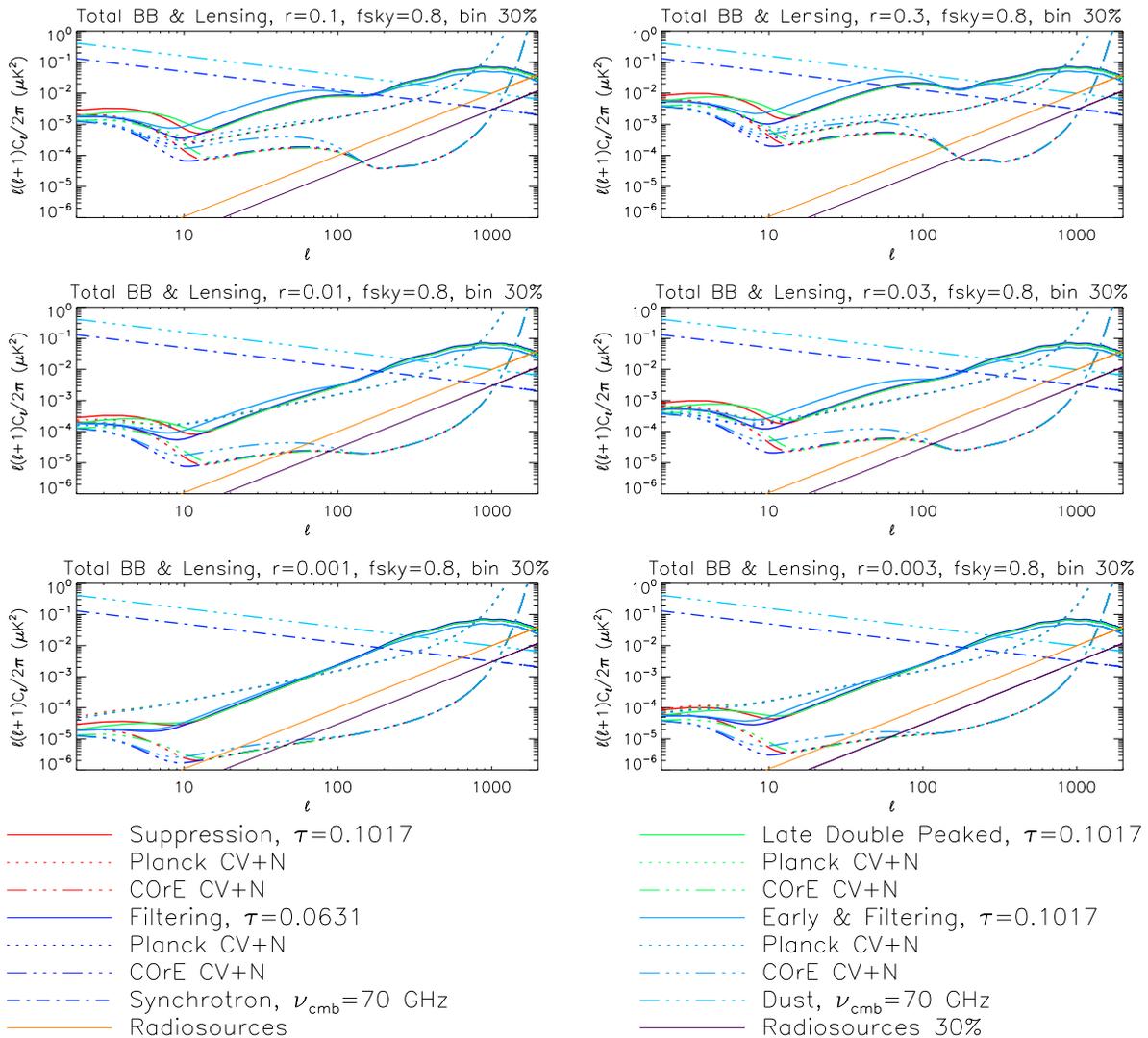}
\vspace*{5.5\baselineskip}
\caption{BB \acs{APS} for the reionization histories: suppression, filtering, late double peaked, early plus filtering for different values of tensor to scalar ratio $r$.
Galactic synchrotron (dash-dotted blue line) and thermal dust (dash-triple-dotted cyan line) polarized emission, and extragalactic point source fluctuations (solid orange line) and their potential residual (solid violet line) as described in previous section (see plot legend and text for details).}
\label{fig:bbstory}
\end{figure}

\begin{figure}[ht]
\centering
\includegraphics[scale=0.9]{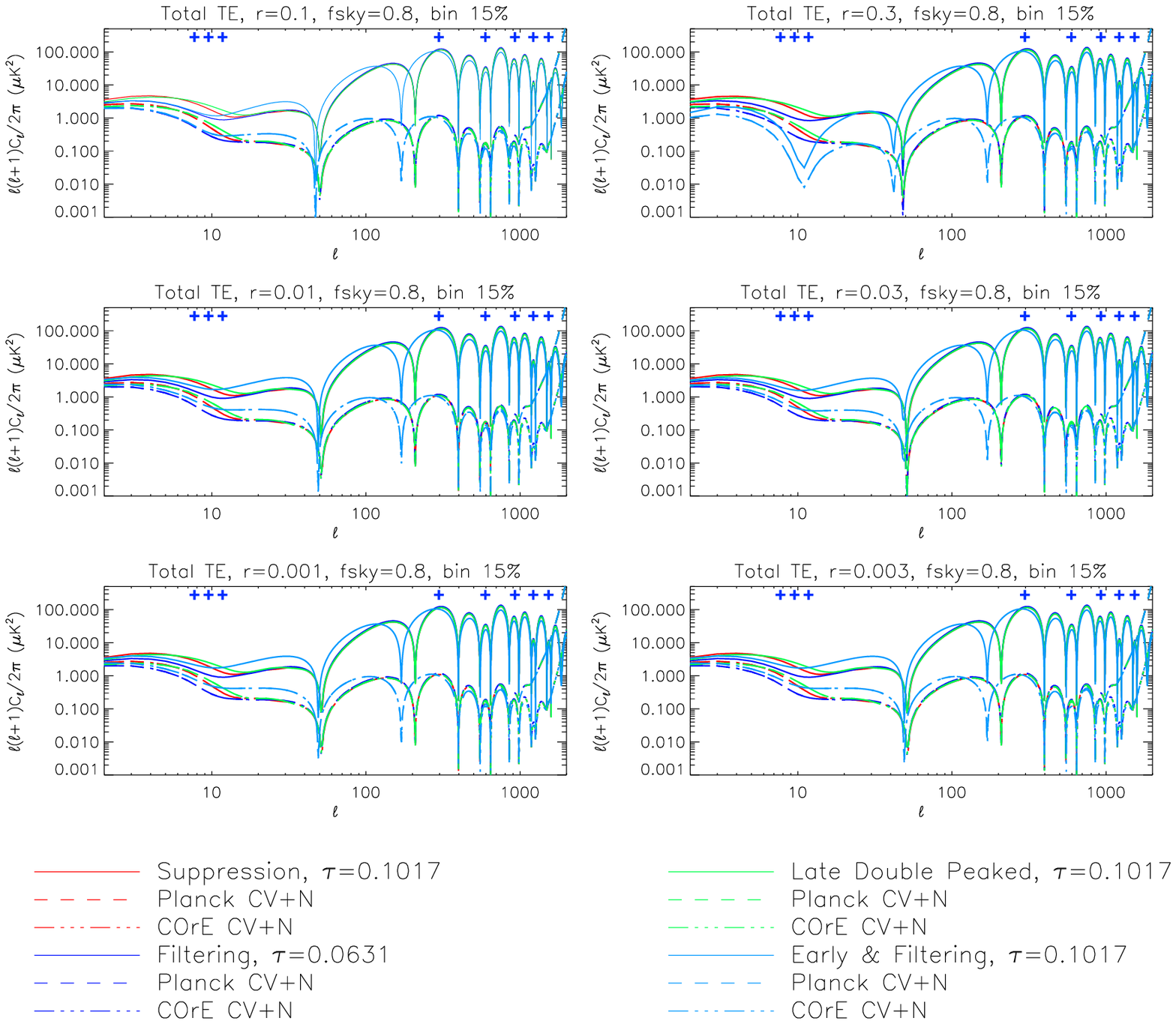}
\vspace*{5.5\baselineskip}
\caption{TE \acs{APS} for the reionization histories: suppression, filtering, late double peaked, early plus filtering (see plot legend and text for details). Here is plotted the absolute value of the TE cross correlation, the ``blue plus'' indicates where the $C_{\ell}$ are positive and each cusp is an inversion point in the sign of the $C_{\ell}$ themselves.}
\label{fig:testory}
\end{figure}

Figs.~\ref{fig:ttaps}, ~\ref{fig:eestory}, ~\ref{fig:bbstory}, ~\ref{fig:testory} show the temperature, polarization and cross-correlation \acs{APS} up to multipoles $\ell = 2000$ for a range of values of the so-called tensor-to-scalar ratio, the $r$ parameter (or $T/S$), which parametrizes the ratio between the amplitudes of primordial tensor and scalar perturbations.
We also show the estimated cosmic and sampling variance, and instrumental noise contributions for the {\it Planck} (dotted lines) and \acs{COrE} (dash-dotted lines) experiments, assuming a sky coverage of $80\%$. \\
\indent As expected, the temperature \acs{APS} does not exhibit a remarkable dependence on the $T/S$ parameter, as evident from the comparison between the panels in Fig.~\ref{fig:ttaps}. Comparing the considered reionization models, we observe a difference between their relative power at high and low multipoles: in particular, the early plus filtering curves show more power at low $\ell$ and less at high $\ell$. Note that this double peaked reionization history has a high redshift phase characterized by a remarkable ionization fraction.

The E-mode power spectrum is shown in Fig.~\ref{fig:eestory}. Again, it is only weakly dependent on the $T/S$ parameter, but for early plus filtering 
model that is widely different from the other models up to the second acoustic peak. Note also that, even if suppression and late (double peaked) models have the same optical depth and cosmological parameters, their power spectra are significantly different in both shape and power. For the late model, the reionization bump is slightly shifted to higher multipoles with respect to the other histories. 

The B-mode power spectrum (see Fig.~\ref{fig:bbstory}), plotted here including also the lensing contribution, 
shows the expected linear dependence on $T/S$ where the primordial signal dominates over the lensing contribution, which, on the contrary, determines the power at high $\ell$, almost independently of $r$.
The tensor-to-scalar ratio has also a strong impact on the shape of the observable acoustic peaks, in particular the first one is flattened for decreasing values of $r$, because of the relative weight of primordial and lensing signal. 
Typically, the reionization bump is stronger for the suppression and the  late (double peaked) models, weaker for the filtering and the early plus filtering histories. 

The ideal sensitivity of {\it Planck} is enough to detect the primordial B-mode for tensor-to-scalar ratios above $few \times 0.01$, in particularly thanks to the information contained in the reionization bump  and up to the first acoustic peak.
The improvement foreseen for an experiment with a sensitivity like \acs{COrE} could allow to reveal the primordial B-mode polarization signal down to $r \simeq 0.001$ (or even lower). 

The ultimate limitation comes from foregrounds. In the case of the B-mode, we show an estimate of the contamination by Galactic synchrotron and thermal dust polarized emission, and by extragalactic point source fluctuations, parametrized as described in the previous section. In all cases, the extragalactic signal is never dominant except at very high multipoles, but still remaining below the contribution by lensing, while Galactic foreground may significantly contaminate the \acs{CMB} measure, especially at low and intermediate multipole.

The set of panels in Fig.~\ref{fig:testory} presents the temperature-polarization cross-correlation \acs{APS}, plotted in absolute value. The plus sign at the top of each panel denote positive correlation. Again, there is only a weak dependence on the $T/S$ parameter, and a substantial difference between the considered evolutionary models, in particular for the early plus filtering history up to $\ell \sim 400$.\\

\subsection{Comparison between models}

In order to understand if different models can be distinguished, we analyze the relative differences between two models and compare them with experimental sensitivity and foreground residual 
estimates (properly normalized).\\
\indent Since we are considering here several models, it is useful to define a reference model to which divide their differences as well as the experimental sensitivity and foreground limitation. We assume here a $\Lambda$\acs{CDM} model with the same cosmological parameters adopted in the models under investigation, and an optical depth fixed by the suppression history ($\tau_{Rif} = \tau_{CF}$). We compute its power spectra using the standard \acs{CAMB}.

Note that, in principle, each model could be normalized in order to match available \acs{CMB} data in the desired range of multipoles. Currently, temperature data play the major role in this respect. 
The overall normalization of the \acs{APS} of a given model is related to the amplitude of initial perturbations, or, equivalently, to the density contrast at a reference scale, typically assumed at $8 \,$h$^{-1} \,$Mpc,
i.e. the parameter $\sigma_{8}$, which is better determined when \acs{CMB} data are combined with galaxy surveys \cite{popa_etal_01}.
In practice, this involves a multiplicative factor of the \acs{APS}. According to this choice, the relative differences between two models could be more or less prominent  at different multipole ranges. In order to make our comparison between models dependent only on the \acs{APS} shape and not on the normalization adopted for each model, before of computing their relative differences, we renormalize each model to make its TT \acs{APS} averaged over $\ell$ equal to that of the reference model.

\begin{figure}[ht]
\centering
\includegraphics[scale=0.9]{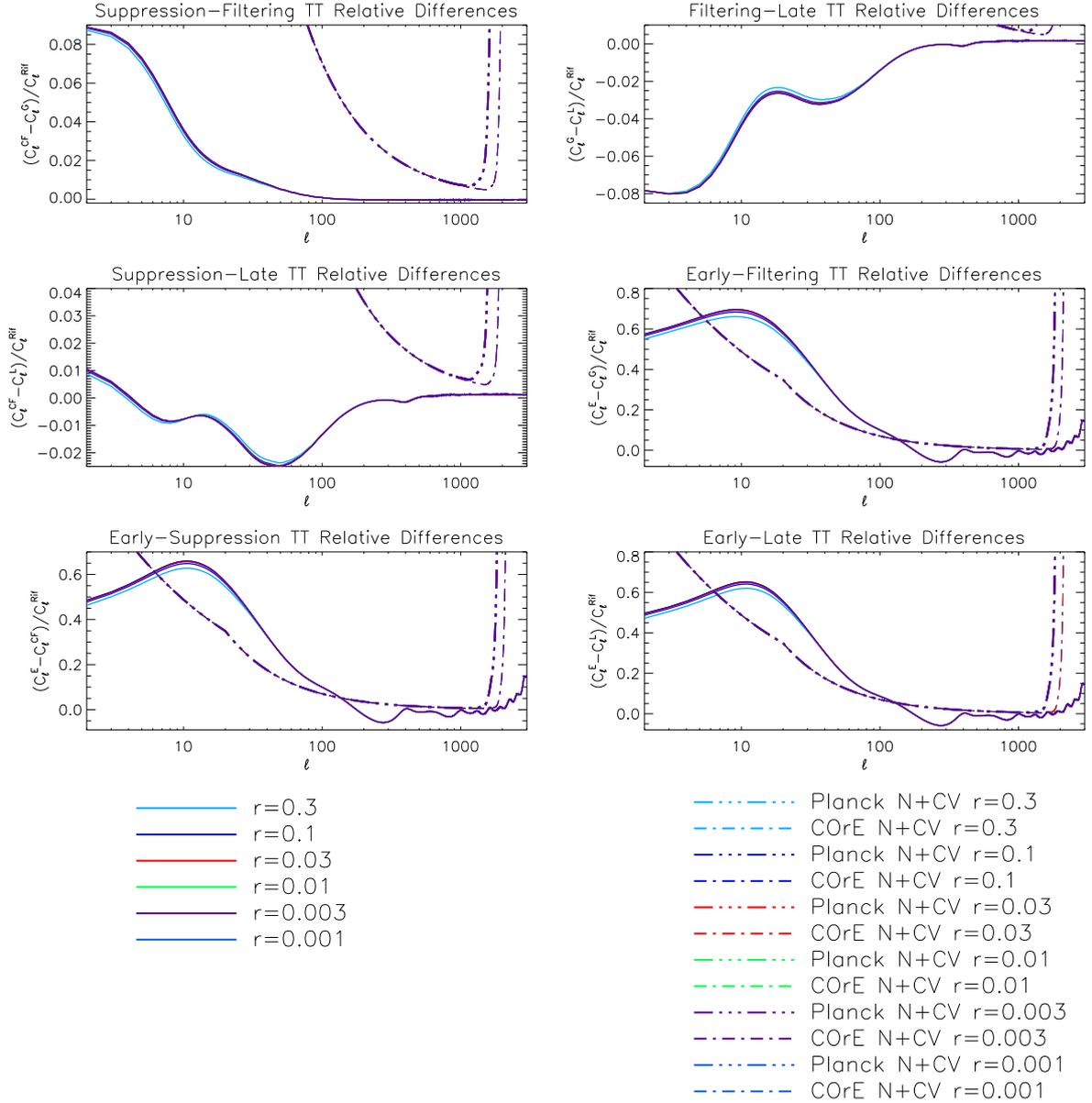}
\vspace*{9.\baselineskip}
\caption{Relative differences between the astrophysical and phenomenological reionization histories in temperature power spectrum for all values of the $r$ parameter assumed in this work (see plot legend and text for details). $C_{\ell}^{Rif}$ is the adopted normalization power spectrum.}
\label{fig:ttdiff}
\end{figure}

\begin{figure}[ht]
\centering
\includegraphics[scale=0.9]{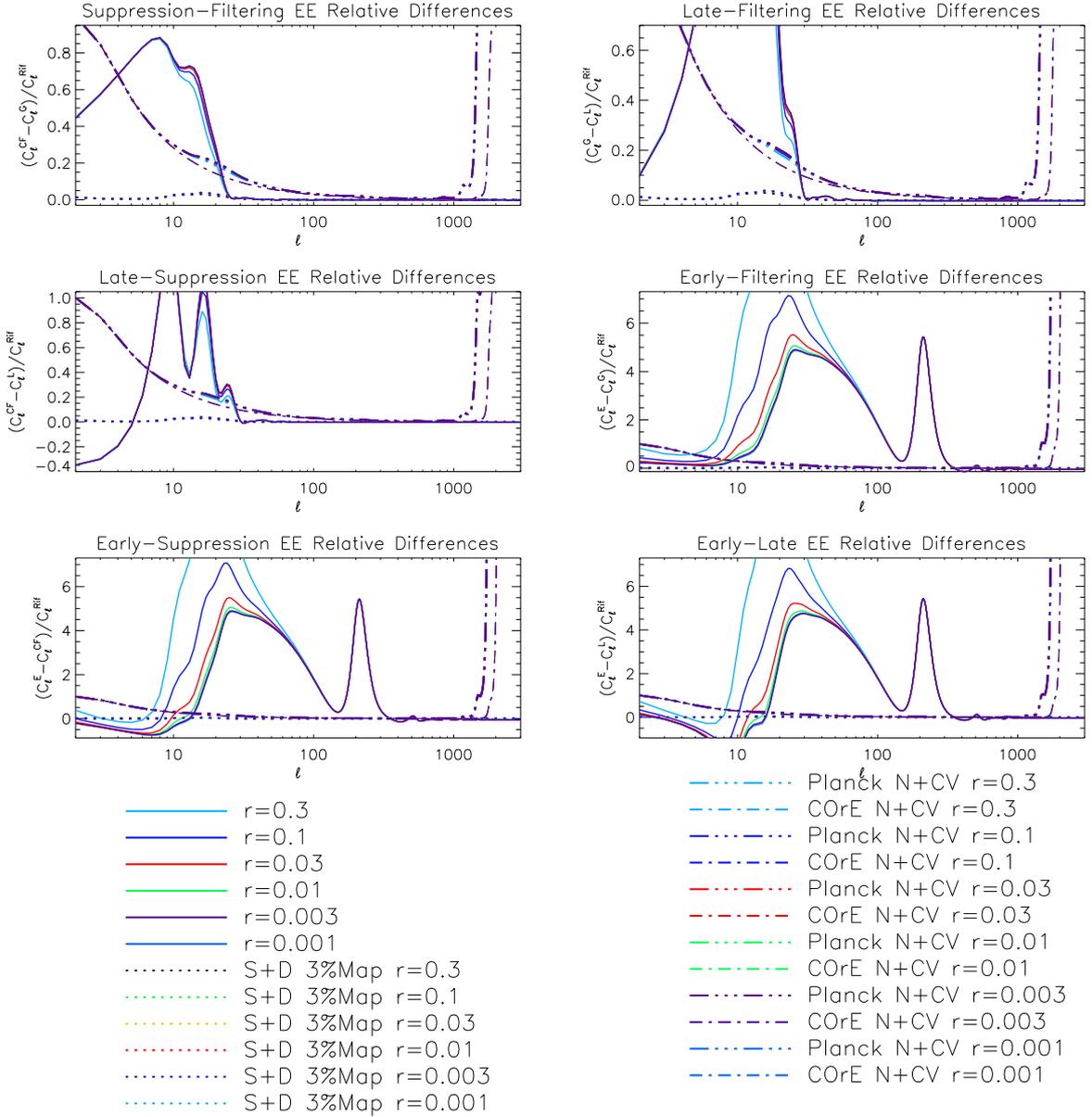}
\vspace*{9.\baselineskip}
\caption{Relative differences between the astrophysical and phenomenological reionization histories in polarization EE-mode power spectrum for all values of the $r$ parameter assumed in this work. $C_{\ell}^{Rif}$ is the adopted normalization power spectrum. Potential residuals of Galactic foregrounds are also shown (see plot legend and text for details). }
\label{fig:eediff}
\end{figure}

\begin{figure}[ht]
\centering
\includegraphics[scale=0.9]{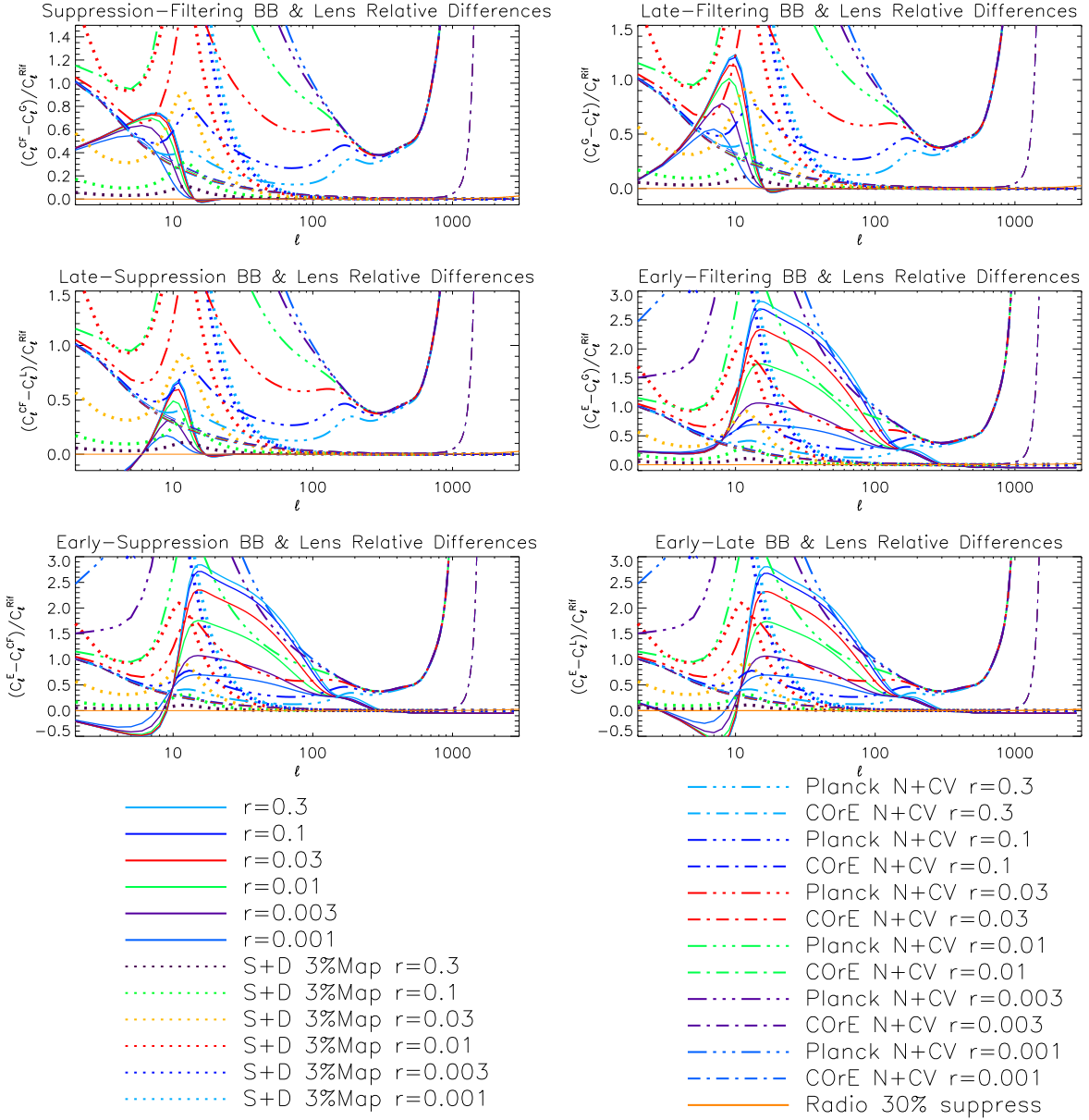}
\vspace*{9\baselineskip}
\caption{Relative differences between the astrophysical and phenomenological reionization histories in polarization B-mode power spectrum for all values of the $r$ parameter assumed in this work. $C_{\ell}^{Rif}$ is the adopted normalization power spectrum. Potential residuals of Galactic foregrounds and extragalactic point source fluctuations are also shown (see plot legend and text for details).}
\label{fig:bbdiff}
\end{figure}

\begin{figure}[ht]
\centering
\includegraphics[scale=0.9]{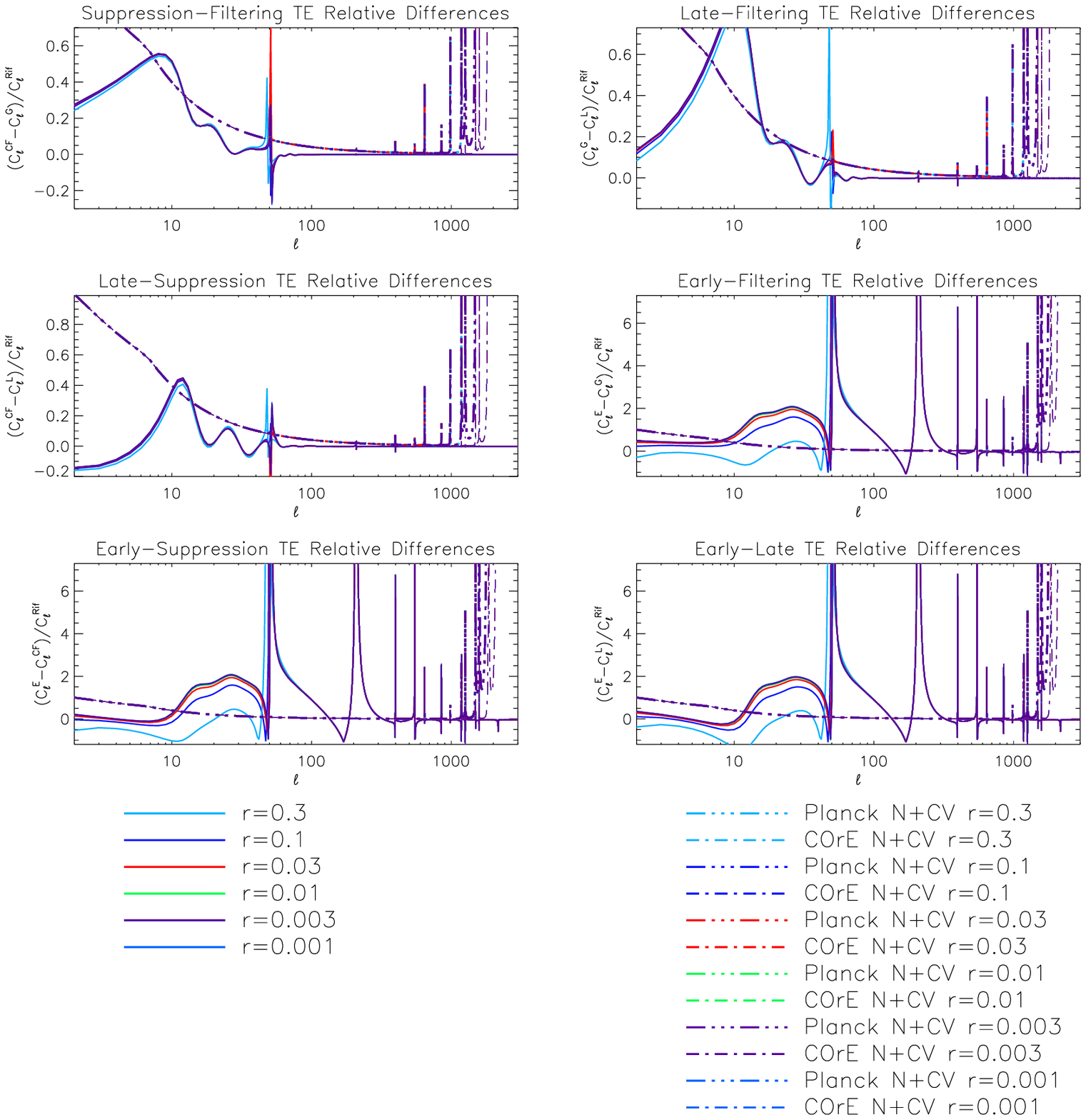}
\vspace*{9.\baselineskip}
\caption{Relative differences between the astrophysical and phenomenological reionization histories in temperature polarization cross correlation power spectrum for all values of the $r$ parameter assumed in this work (see plot legend and text for details). 
$C_{\ell}^{Rif}$ is the adopted normalization power spectrum. } 
\label{fig:tediff}
\end{figure}

We report in Figs.~\ref{fig:ttdiff}, ~\ref{fig:eediff}, ~\ref{fig:bbdiff}, ~\ref{fig:tediff} the relative differences between models for the temperature anisotropies, polarization EE and BB \acs{APS}, and the temperature-polarization cross-correlation, respectively. Each panel shows the comparison between two models, normalized to the reference model defined above, for a wide set of the $T/S$ parameter.

As anticipated, the tensor-to-scalar ratio does not affect significantly the temperature anisotropies, and for this reason in Fig.~\ref{fig:ttdiff} the curves appear almost superimposed. \\
In addition, at $\ell \gsim 100-300$ the differences tend to be approximately null, except for the comparison of the early plus filtering model with all the others, because of their dramatic difference at high redshifts. \\
Considering the whole multipole range, the largest differences appear again in the comparison of the early plus filtering model with all the other (later) histories, achieving a maximum level of $\sim 70\%$ at $\ell \sim 10-15$. \\
Note that we can discriminate only early processes from other (later) models using only the \acs{CMB} TT \acs{APS}. This can not be significantly improved with the future generation of experiments, the limitation coming essentially from cosmic variance. Thus, polarization data are crucial. 

The comparison between the E-mode polarization power spectra is more interesting (see Fig.~\ref{fig:eediff}).
As evident, in particular in the three last panels in Fig.~\ref{fig:eediff} where the early plus filtering scenario is compared with the others, the differences related to the $r$ parameter emerge clearly.
The greater is $r$ the greater are the relative differences. The large difference at $\ell \sim 200$ is due to the first early reionization phase, absent in the other models. \\
\indent In general, remarkable relative differences appear between all the models at $\ell \sim 5-25$ or even up to $\ell \sim 100$ when comparing the early plus filtering scenario with the others. They are connected to the details of the ionization history at lower redshifts.
We also plot a potential residual contamination from Galactic foregrounds. Note that a foreground removal at a few per cent level of accuracy at map level makes astrophysical contamination below the sensitivity limitation of on-going and future space experiments. The difference in nominal capability of {\it Planck} and \acs{COrE} is not so remarkable in this respect, although, in practice, a significant improvement in sensitivity and frequency coverage clearly will make much more robust and accurate the foreground subtraction process.

More complex is the case of the B-mode polarization power spectra, reported in Fig.~\ref{fig:bbdiff}, where, for completeness, we display also an estimate of the potential residual from extragalactic point source fluctuations which is always negligible in comparison with the other sources of limitation.
As before, the early plus filtering model largely differ from the others. Note that for the B-mode the relative differences between the various models remarkably depend on $r$ in all cases. 
In this representation, the {\it Planck} sensitivity strikingly depends on $r$ at low and intermediate $\ell$, because of the instrumental noise limitation, while that of \acs{COrE} is almost independent on $r$, being essentially cosmic variance limited where the lensing contribution does not dominate.
The sensitivity of \acs{COrE} instruments is such that we could be able to distinguish between suppression and filtering (or late) models in a certain range of multipoles (around $\ell \sim 10$) even for $r$ slightly larger 
than $\sim 10^{-3}$, while the early plus filtering history can be distinguished from all the other histories on a wider multipole region (or for even lower values of $r$).
The impact of residuals of Galactic polarized foregrounds clearly increases for decreasing tensor-to-scalar ratio, as expected in the comparison of relative differences between models. Only for $r \gsim 0.1$ a foreground subtraction at 3\% accuracy level in the map is enough to discriminate between each couple of the considered models, while the early plus filtering history can be distinguished from all the other histories even with a less accurate foreground subtraction, thanks to its prominent differences at intermediate multipoles.

Finally, relative differences in the temperature-polarization cross-correlation \acs{APS} are also very large on very wide ranges of multipole when comparing the early plus filtering history with all the other models 
(see Fig.~\ref{fig:tediff}), filtering and late models show remarkable differences (larger than $80\%$) at low multipoles, while the suppression model differ from the late and filtering models only on a very small range of multipoles  around $\ell \sim 10$.
The ``spikes`` appearing at high multipoles are due to little shifts of the multipoles corresponding to the change of sign of the cross-correlation spectra for the different considered models. 
The difference in nominal capability of {\it Planck} and \acs{COrE} is not so remarkable for the TE mode.

\section{Conclusion}
\label{conclu}

The inclusion of astrophysically motivated ionization and thermal histories in numerical codes is crucial for the accurate prediction of the features induced in the \acs{CMB}, for constraining reionization 
models with \acs{CMB} data, and to exploit current and future high quality \acs{CMB} data with great versatility to accurately extract cosmological information.

We have implemented a modified version of \acs{CAMB}, the Cosmological Boltzmann code for computing the \acs{APS} of the anisotropies of the \acs{CMB}, 
to introduce the hydrogen and helium ionization fractions predicted in two astrophysical reionization models, i.e. suppression and filtering model, in two classes of phenomenological reionization histories, involving late or early reionization, and in their combination, as alternative to the original implementation of reionization in the \acs{CAMB} code, and beyond the simple $\tau$-parametrization.
For astrophysical models, we provide also suitable analytical descriptions of the ionization and thermal histories that can be ingested in any numerical code aimed at computing \acs{CMB} features.
We compared the results obtained for these models for all the non-vanishing (in the assumed scenarios) modes of the \acs{CMB} \acs{APS}.
As a  significant step forward with respect to previous analyses, the emphasis has been posed here to the extension to a first detailed characterization of the polarization B-mode \acs{APS}. 
Its amplitude and shape depends, in particular, on the tensor-to-scalar ratio, $r$, and on the reionization history, thus an accurate modeling of the reionization process will have implications for the precise determination 
of $r$ or to set more precise constraints on it through the joint analysis of E and B-mode polarization data available in the next future and from a mission of next generation. 
This is particularly  crucial in the case of low values of $r$, because of the contribution to B-mode coming from lensing that competes with the primordial B-mode at intermediate multipoles.
Taking into account also the limitation from potential residuals of astrophysical foregrounds, we discussed the capability of next data to disentangle between different reionization scenarios in a wide range of tensor-to-scalar ratios.  

\section*{Acknowledgements}

\noindent
We acknowledge the use of the Legacy Archive for Microwave Background Data
Analysis (LAMBDA). Support for LAMBDA is provided by the NASA Office of
Space Science.
We acknowledge support by ASI through ASI/INAF Agreement I/072/09/0 for
the {\it Planck} LFI Activity of Phase E2 and by MIUR through PRIN 2009.
 
\appendix
\appendixpage

\section{Fitting ionization and thermal histories}
\label{fitting}

The considered astrophysical reionization models, as well as others published in the literature,
provide ionization and thermal histories in tabulated form. 
The cosmological analysis of \acs{CMB} anisotropy data largely relies on Boltzmann codes 
for computing the angular power spectrum in temperature, polarization, and cross-correlation
modes under general conditions. The inclusion in such codes of reionization histories beyond the 
simplistic phenomenological approximations already implemented in the publicly available codes,
allows to achieve more accurate predictions, of particular interest for the analysis of polarization data. 
Having functional descriptions of the evolution of the 
ionization fraction allows to speed-up computation and improve code versatility with respect to the use of interpolation of tabulated grids. 
Although in this context only the ionization history is relevant, we report here for completeness also the results concerning the thermal history.

The software we used to fit to the suppression and filtering model is \emph{Igor Pro (v. 6.21)}, 
an integrated program for visualizing, analyzing, transforming and presenting experimental data, such as curve-fitting, Fourier transforms, smoothing, statistics, and other data analysis, image display and processing, by a combination of graphical and command-line user interface. 
We report here all the functional forms we found particularly suitable to represent the considered ionization and thermal histories, since they could be useful as guidelines to fit other kinds of reionization history.
More details about the usage of this software to the current aims and the complete list of parameters of the below functions are given in \cite{trombetti_burigana_rep589}.

\subsection{Fitting functions for the reionization histories}
\label{reionfit}

In the case of the suppression model the redshift intervals and the relative analytic functions are: 
           
- Interval $z={[0,3.8]}$ - Polinomial Function of $6^\circ$ order:
	
           $$\chi_{re}=a_0+a_1z+a_2z^2+a_3z^3+a_4z^4+a_5z^5+a_6z^6.$$\\
           
- Interval $z={[3.8,6]}$ -  Polinomial Function of $5^\circ$ order:

           $$\chi_{re}=b_0+b_1z+b_2z^2+b_3z^3+b_4z^4+b_5z^5.$$\\
           
- Interval $z={[6,9]}$ -  Polinomial Function of $5^\circ$ order:     

           $$\chi_{re}=c_0+c_1z+c_2z^2+c_3z^3+c_4z^4+c_5z^5.$$\\

- Interval $z=[9,12.4]$ - Log-Normal Function:  

           $$\chi_{re}=l_0 + l_1 {\rm exp}   \left[{  - \left({   \frac {{\rm ln} (z/l_2)}  {l_3}  }\right)^2 }\right] .$$\\

- Interval $z=[12.4,14.2]$ - Sigmoidal Function: 

           $$\chi_{re}=d_0 + \frac{d_1}{1 + {\rm exp} \left( \frac{d_2 - z}{d_3}\right )}.$$\\

- Interval $z=[14.2,16.8]$ - Hill Equation:  

           $$\chi_{re}=e_0 + (e_1 - e_0)\frac{z^{e_2}}{z^{e_2} + e_3^{e_2}}.$$\\

- Interval $z=[16.8,18]$ - Decaying Exponential Function (expXOffset):

           $$\chi_{re}=h_0 + h_1 {\rm exp} \left( \frac{h_3 - z}{h_2} \right).$$\\

- Interval $z=[18,20]$ - Hill Equation:

           $$\chi_{re}=i_0 + (i_1 - i_0)\frac{z^{i_2}}{z^{i_2} + i_3^{i_2}}.$$\\

- Interval $z=[20,20.2]$ - Linear Function:

           $$\chi_{re}=m_0 + m_1z.$$\\

- Interval $z=[20.2,23]$ - Decaying Exponential Function (expXOffset):
           
           $$\chi_{re}=f_0 + f_1 {\rm exp} \left( \frac{f_3 - z}{f_2} \right).$$\\

- Interval $z=[23,30]$ - Log-Normal Function:

           $$\chi_{re}=g_0 + g_1 {\rm exp}  \left[{  - \left({   \frac {{\rm ln} (z/g_2)}  {g_3}  }\right)^2 }\right].$$\\
           
 When considering the filtering model the fitting functions are:

- Interval $z=[0,3.8]$ - Polinomial Function of  $9^\circ$ order:
           
           $$\chi_{re}=a_0+a_1z+a_2z^2+a_3z^3+a_4z^4+a_5z^5+a_6z^6+a_7z^7+a_8z^8+a_9z^9.$$\\

- Interval $z=[3.8,6]$ - Log-Normal Function:
           
           $$\chi_{re}=i_0 + i_1  {\rm exp}  \left[{  - \left({  \frac {{\rm ln} (z/i_2)}  {i_3}  }\right)^2 }\right].$$\\

- Interval $z=[6,6.2]$ - Linear Function:

           $$\chi_{re}=m_0 + m_1z.$$\\

- Interval $z=[6.2,9]$ - Hill Equation:

           $$\chi_{re}=l_0 + (l_1 - l_0) \frac{z^{l_2}}{z^{l_2} + l_3^{l_2}}.$$\\

- Interval $z=[9,11.6]$ - Sigmoidal Function:
           
           $$\chi_{re}=b_0 + \frac{b_1}{1 + {\rm exp} \left( \frac{b_2 - z}{b_3} \right)}.$$\\

- Interval $z=[11.6,13]$ - Power Function:

           $$\chi_{re}=c_0 + c_1z^{c_2}.$$\\

- Interval $z=[13,15]$ - Power Function:

           $$\chi_{re}=h_0 + h_1z^{h_2}.$$\\

- Interval $z=[15,17]$ - Decaying Exponential Function (expXOffset):
           
           $$\chi_{re}=d_0 + d_1 {\rm exp} \left( \frac{d_3 - z}{d_2} \right).$$\\

- Interval $z=[17,19.6]$ - Decaying Exponential Function (expXOffset):
           
           $$\chi_{re}=e_0 + e_1 {\rm exp} \left( \frac{e_3 - z}{e_2} \right).$$\\

- Interval $z=[19.6,22.2]$ - Decaying Exponential Function (expXOffset):
           
           $$\chi_{re}=f_0 + f_1 {\rm exp} \left( \frac{f_3 - z}{f_2} \right).$$\\

- Interval $z=[22.2,30]$ - Decaying Exponential Function (expXOffset):

           $$\chi_{re}=g_0 + g_1 {\rm exp} \left( \frac{g_3 - z}{g_2} \right).$$\\

\subsection{Fitting functions for the temperature histories}
\label{tempfit}

In the case of the electron temperature, the fitting functions for the suppression model are:

- Interval $z=[0,3.8]$ - Polinomial Function of $8^\circ$ order:
           
           $$T_{re}=a_0+a_1z+a_2z^2+a_3z^3+a_4z^4+a_5z^5+a_6z^6+a_7z^7+a_8z^8.$$\\

- Interval $z=[3.8,5.8]$ - Sigmoidal Function:
           
           $$T_{re}=b_0 + \frac{b_1}{1 + {\rm exp} \left( \frac{b_2 - z}{b_3} \right)}.$$\\

- Interval $z=[5.8,13]$ - Hill Equation:
           
           $$T_{re}=c_0 + (c_1 - c_0) \frac{z^{c_2}}{z^{c_2} + c_3^{c_2}}.$$\\

- Interval $z=[13,16]$ - Hill Equation:
          
          $$T_{re}=l_0 + (l_1 - l_0) \frac{z^{l_2}}{z^{l_2} + l_3^{l_2}}.$$\\

- Interval $z=[16,16.8]$ - Decaying Exponential Function (expXOffset):
           
           $$T_{re}=d_0 + d_1 {\rm exp} \left( \frac{d_3 - z}{d_2} \right).$$\\

- Interval $z=[16.8,18.6]$ - Hill Equation:

           $$T_{re}=e_0 + (e_1 - e_0) \frac{z^{e_2}}{z^{e_2} + e_3^{e_2}}.$$\\

- Interval $z=[18.6,20.2]$ - Log-Normal Function:

           $$T_{re}=h_0+h_1  {\rm exp}  \left[{  - \left({  \frac {{\rm ln} (z/h_2)}  {h_3}  }\right)^2 }\right].$$\\

- Interval $z=[20.2,21]$ - Sigmoidal Function:
           
           $$T_{re}=i_0 + \frac{i_1}{1 + {\rm exp} \left( \frac{i_2 - z}{i_3} \right)}.$$\\

- Interval $z=[21,22]$ - Log-Normal Function:

           $$T_{re}=g_0 + g_1  {\rm exp}  \left[{  - \left({  \frac {{\rm ln} (z/g_2)}  {g_3}  }\right)^2 }\right].$$\\

- Interval $z=[22,30]$ - Log-Normal Function:

           $$T_{re}=f_0 + f_1  {\rm exp}  \left[{  - \left({  \frac {{\rm ln} (z/f_2)}  {f_3}  }\right)^2 }\right].$$\\

Finally, the fitting functions for the filtering model are:\\

- Interval $z=[0,3.8]$ - Polinomial Function of $8^\circ$ order:
           
           $$T_{re}=a_0+a_1z+a_2z^2+a_3z^3+a_4z^4+a_5z^5+a_6z^6+a_7z^7+a_8z^8.$$\\

- Interval $z=[3.8,6.2]$ - Sigmoidal Function:
          
           $$T_{re}=b_0 + \frac{b_1}{1 + {\rm exp} \left( \frac{b_2-z}{b_3} \right) }.$$\\

- Interval $z=[6.2,10]$ - Log-Normal Function:
           
           $$T_{re}=c_0 + c_1  {\rm exp}  \left[{  - \left({  \frac {{\rm ln} (z/c_2)}  {c_3}  }\right)^2 }\right].$$\\
           
- Interval $z=[10,11]$ - Decaying Exponential Function (expXOffset):
          
           $$T_{re}=h_0 + h_1 {\rm exp} \left( \frac{h_3 - z}{h_2} \right) .$$\\

- Interval $z=[11,15]$ - Sigmoidal Function:
           
           $$T_{re}=d_0 + \frac{d_1}{1 + {\rm exp} \left( \frac{d_2 - z}{d_3} \right) }.$$\\

- Interval $z=[15,18]$ - Sigmoidal Function:
           
           $$T_{re}=e_0 + \frac{e_1}{1 + {\rm exp} \left( \frac{e_2 - z}{e_3} \right) }.$$\\

- Interval $z=[18,21]$ - Log-Normal Function:
          
           $$T_{re}=f_0 + f_1  {\rm exp}  \left[{  - \left({  \frac {{\rm ln} (z/f_2)}  {f_3}  }\right)^2 }\right ] .$$\\

- Interval $z=[21,30]$ - Log-Normal Function:
          
           $$T_{re}=g_0 + g_1  {\rm exp}  \left[{  - \left({   \frac {{\rm ln} (z/g_2)}  {g_3}  }\right)^2 }\right ] .$$

\section{Code implementation in \acs{CAMB}}
\label{codecamb}

We give here some details on the routines we implemented in \acs{CAMB} to make it able to reproduce the considered histories for the ionization fraction and electron temperature, summarizing the main innovations included in the modules of interest.

\subsection{Subroutine modifications in Reionization module}

The first improvement concern the type {\it ReionizationParams} with the inclusion of a new string variable, {\it history}, through which the user can discriminate between the models, stored in the settings parameters file {\it params.ini}. 

The main function of the original module, {\it $Reionization\_{xe}$}, has been now written for each history, {\it $Reionization\_{xeCF}$}, {\it $Reionization\_{xeG}$}, {\it $Reionization\_{xeL}$}, {\it $Reionization\_{xeE}$}. All of them retrieve the analytical ionization fraction for each redshift bin of interest. \\ 
With this approach is not necessary to parametrize $\chi_{e}$ in terms of the variable {\it WindowVarMid}:

\begin{equation}
y = (1+z)^{3/2} \, , 
\end{equation}

\noindent
where the exponent is set with the constant {\it Reionization\_zexp}.

Some initial parameter values have been revisited too, such as the maximum redshift at which $\chi_{e}$ varies, fixed to $700$ instead of $40$ to take into account high redshift reionization phases, necessary for the early history, and the corresponding initial scale factor {\it astart}, inversely proportional to the redshift:

\begin{equation}
a=\frac{1}{1+z} \, .   
\end{equation}

In the function {\it $Reionization\_{timesteps}$} the minimum number of time steps between {\it $tau\_{start}$} and {\it $tau\_{complete}$}, the relevant times for the reionization process, has been incremented to  $1000$, while for the implementation of the adopted ionization histories  the functions listed below are no longer necessary:\\ 

\noindent
- {\it $Reionization\_{doptdepth(z)}$}, the subroutine that expresses the integral optical depth in terms of the scale factor, \\ 
- {\it $Reionization\_GetOptDepth(Reion, ReionHist)$}, the routine which evaluates the integral of the optical depth in the redshift interval $(0,z_{max}^{reion})$, \\ 
- {\it $Reionization\_{zreFromOptDepth(Reion, ReionHist)}$}, a general routine to find the $z_{re}$ parameter given optical depth, \\
- {\it $Reionization\_{SetFromOptDepth(Reion, ReionHist)}$}, the subroutine that calculates the redshift of reionization. \\ 

\noindent
This set of function, in fact, is related to the optical depth definition, i.e.

\begin{equation}
\tau = \int_{0}^{\eta_{0}} d\eta a n_{e} \sigma_{T} \, ,  
\end{equation}
\\
\noindent
as computed in the standard \acs{CAMB}, while for the histories under examination we have implemented specific codes to evaluate it, both in our modified \acs{CAMB} version and as independent codes.
 
\subsection{Subroutine modifications in ThermoData module}

The {\it Thermodata} module, implemented in {\it modules.f90} source file, contains the subroutine {\it inithermo(taumin, taumax)}, which evaluates 
the unperturbed baryon temperature and ionization fraction as function of time. If there is reionization, the function discriminates between the models, 
smoothly increases $\chi_{e}$ to the requested value and sets the {\it $actual\_opt\_depth$} to the value imposed by the corresponding model.

\vskip 0.5cm


\end{document}